\documentclass[12pt]{article}
\usepackage[T1]{fontenc}
\usepackage[utf8]{inputenc}
\usepackage[brazil,english]{babel}
\usepackage{amssymb,amsmath,amsbsy}
\usepackage[colorlinks=true]{hyperref}
\usepackage{lscape}
\usepackage{graphicx}
    \graphicspath{{./Figures/}}
\usepackage{subfigure}
\usepackage{floatrow}
\usepackage{mathtools}
\usepackage[tmargin=3cm,lmargin=3cm,rmargin=2cm,bmargin=2cm]{geometry}
\usepackage{setspace}
\usepackage[square,numbers,sort&compress]{natbib}
\usepackage{longtable}
\usepackage{xcolor}
\usepackage{microtype}
\usepackage{multirow}
\usepackage{stix}
\usepackage{array}
\usepackage{caption}

\DeclareMathOperator{\sech}{sech}
\newcommand{\sqabs}[1]{{\left|#1\right|}^2}
\newcommand{\ZZ}{\mathbb{Z}}
\newcommand{\RR}{\mathbb{R}}
\newcommand{\CC}{\mathbb{C}}
\newcommand{\II}{\mathbb{I}}
\newcommand{\hypergeom}[1]{{}_2F_1\!\left(#1\right)}
\newcommand{\diff}[2]{\frac{\text{d}#1}{\text{d}#2}}
\newcommand{\ddiff}[2]{\frac{\text{d}^2#1}{{\text{d}#2}^2}}

\title{{The kinetic Hamiltonian with position-dependent mass}}
\author{ R. M. Lima$^{1}$ \ {\normalsize{and}}\  H. R. Christiansen$^{1,2}\footnote{Corresponding author. Electronic address: hugo.christiansen@ifce.edu.br}$ 
  \\
 \footnotesize{IFCE -\ Instituto Federal de Educação, Ciência e Tecnologia do Ceará,} \\
 \footnotesize{CEP 62042-030 - Sobral$^{1}$, and CEP 61940-750 - Maranguape$^{2}$, Ceará, Brazil}\\ 
   }

\date{}

\begin{document}

\maketitle	

\begin{abstract}
In the present paper we examine in a systematic way the most relevant orderings of {\it pure kinetic Hamiltonians} for five different position-dependent mass (PDM) profiles:  {soliton-like}, reciprocal quadratic and {biquadratic}, exponential and parabolic. As a result of the non-commutativity between momentum and position operators, a diversity of effective potentials is generated. We analyse the whole set and find unexpected coincidences as well as discrepancies among them. We obtain analytically the full-spectrum of energies and solutions in the twenty-five cases considered. It is shown how the simple ordinary constant-mass solutions are transformed into a variety of complex combinations of transcendental functions and arguments. We find that particles with a non-uniform mass density can present discrete energy spectra as well as continuous ones which can be bounded or not. These results are consistent with the fact that although the external potential is zero, PDM eigenfunctions are not actual free states but a sort of effective waves in a solid-state sample. This is precisely the origin of the position-dependent mass. In all the events we obtain exact complete spectral expressions. Our methodological procedure thus puts a wide diversity of Hamiltonian seeds on an equal footing in order to be compared. This allows choosing the better arrangement to model a specific solid or heterostructure once the spectrum of a given material is experimentally available.   {Finally, we perform a one-dimensional model calculation of a double heterostructure with a parabolic PDM particle in the interface region}. Our study is also indicated for applications inside material structures with the addition of external potentials.
\end{abstract}

\\

{\small Published in \textit{Physica E: Low dimensional Systems and Nanostructures 150 (2023) 115688.}}

%%%%%
%% Section 1 Introduction
%%%%%

\section{Introduction}

Quantum Mechanics of particles with non-uniform masses has always been matter of concern for the Hamiltonian's dependence on the ordering of the dynamical variables. This problem was born in the realm of solid-state physics \cite{wannier:1937,slater:1949,luttinger:kohn:1955,bendaniel:duke:1966,gora:williams:1969} and soon became of interest from a rather more fundamental point of view regarding basic symmetries \cite{shewell:1959,vonroos:1983,thomsen:einevoll:hemmer:1989,levy-leblond:1995,chetouani:dekar:hammann:1995}.

Position-dependent mass (PDM) models have been under regular development since the sixties and continuously improved up to present days. Particularly, in the last decade modeling quantum systems with PDM particles has grown as a consequence of its wide area of applications \cite{cunha:christiansen:2013,christiansen:cunha:2013,christiansen:cunha:2014,dacosta:gomez:portesi:2020,ho:roy:2019,schmidt:dejesus:2018}. Among them, this technique has been applied to the understanding of the electronic properties of semiconductor heterostructures, crystal-growth techniques \cite{gora:williams:1969,bastard:furdyna:mycielski:1975}, quantum wells and quantum dots \cite{harrison:valavanis:2016,serra:lipparini:1997,burt:1992,alhaidari:2002,alhaidari:2003,yu:dong:2004,dong:lozada-cassou:2005,mustafa:mazharimousavi:2007,lima:etal:2012,elnabulsi:2020,elnabulsi:2020b}, helium clusters \cite{barranco:etal:1997}, graded crystals \cite{geller:kohn:1993}, quantum liquids \cite{desaavedra:etal:1994}, nanowire structures with size variations, impurities, dislocations, and geometry imperfections \cite{willatzen:lassen:2007,peter:2008,elnabulsi:2020c,christiansen:lima:2023}, as well as in superconductors investigations \cite{bendaniel:duke:1966,gora:williams:1969,bastard:furdyna:mycielski:1975,bastard:1981,zhu:kroemer:1983,vonroos:1983,bastard:1988,galbraith:duggan:1988,geller:kohn:1993,ren:chang:2010}.

Concomitantly, the dynamics of quantum systems has been extensively studied in connection with all kind of problems, from low to high energy physics. Mathematically, the problem consists in dealing with a Schrodinger equation, which, for stationary systems, results in an eigenvalue problem in a Hilbert space \cite{cohen-tannoudji:diu:laloe:1977}.

The number of exactly solvable potentials is relatively limited in ordinary quantum mechanics but as soon as we switch the constant mass into a position-dependent distribution the mathematical challenge grows dramatically. In order to treat analytically these problems, different methods have been used. We will apply point canonical transformations, which keep invariant the canonical form of the wave equation \cite{dekar:chetouani:hammann:1998,dekar:chetouani:hammann:1999,alhaidari:2002,alhaidari:2003,cunha:christiansen:2013}, but there exist alternative procedures like super-symmetric quantum mechanics \cite{plastino:etal:1999,gonul:etal:2002,karthiga:ruby:senthilvelan:2018}, Lie algebras \cite{roy:roy:2002} and Weyl integral quantization  \cite{gazeau:2020, gazeau:2022} 
(more refs. in \cite{yu:dong:2004,rajbongshi:2018}), which have been also used in the last years.

For a non-uniform mass, \(M\equiv M(\boldsymbol{r})\), the most general Hermitian kinetic Hamiltonians
\footnote{ {
For a digression on non-Hermitian Hamiltonians see \cite{jiang:yi:jia:2005,mustafa:mazharimousavi:2006}. 
It is also worth noting that the parity-time symmetry, although weaker than hemiticity, has been proven to allow unitary time-evolution in some cases; see \cite{bender:2015} and \cite{elnaga:2019}.}
} 
is given by \cite{vonroos:1983}
\begin{equation}
    \label{eq:1-1}
    \hat{T}_{ {a,b}}(\boldsymbol{r})=\frac{1}{4}\left(M^a\hat{\boldsymbol{p}}M^{-1-a-b}\hat{\boldsymbol{p}}M^b+M^b\hat{\boldsymbol{p}}M^{-1-a-b}\hat{\boldsymbol{p}}M^a\right)\;,
\end{equation}
where \(a,b\in\RR\) are mass parameters and \(\hat{\boldsymbol{p}}\equiv-i\hbar\boldsymbol{\nabla}\).

Some particular cases of this expression are common in the literature and will be here studied and abbreviated by BDD  {(BenDaniel \& Duke)} \cite{bendaniel:duke:1966} ($a=b=0$), GW  {(Gora \& Williams)} \cite{gora:williams:1969,bastard:furdyna:mycielski:1975,bastard:1981,li:kuhn:1993} ($a=-1$, $b=0$), ZK  {(Zhu \& Kroemer)} \cite{zhu:kroemer:1983} ($a=b=-1/2$), LK  {(Li \& Kuhn)} \cite{li:kuhn:1993} ($a=0$, $b=-1/2$) and MM  {(Mustafa \& Mazharimousavi)} \cite{mustafa:mazharimousavi:2007} ($a=b=-1/4$) Hamiltonians (or orderings). In \cite{vonroos:1983} the inherent ambiguity resulting from the mass parameters and the breaking of Galilean invariance (except when considering the BDD ordering\footnote{Indeed, in \cite{levy-leblond:1995} BDD is deduced after Galilean invariance (similar results can be found in \cite{thomsen:einevoll:hemmer:1989}).}) are also pointed out --- however, it is worth noting that other authors \cite{lima:etal:2012,gonul:etal:2002} came up with generalizations and even alternatives to the von Roos' Hamiltonian shown to respect Galilean transformations ---; furthermore, when abrupt heterojunctions (characterized by finite discontinuities in the position dependent mass) are considered \cite{morrow:brownstein:1984} just $a=b$ orderings are viable, provided the continuity conditions (in one dimension) are imposed on \(M^a\Psi(x)\) and \(M^{-1- {a}}\Psi'(x)\). 
  {The manisfest Galilean invariance and the vanishing of the ambiguous potential of kinetic origin (see below) seem to be the main reasons that made of BDD the most consensual kinetic Hamiltonian in the literature e.g. \cite{dekar:chetouani:hammann:1998,dekar:chetouani:hammann:1999,alhaidari:2002,alhaidari:2003}. Some experimental fits with BDD mass parameters can be also found in \cite{csavinszky:elabsy:1988,galbraith:duggan:1988}}. Notwithstanding, the other orderings have been adopted with success in some systems \cite{rajbongshi:2018}, particularly when interfaces are not brusque. For example, in the case of exponential modeling of potential and mass within a material, the BDD ordering is the one to be discarded \cite{desouzadutra:almeida:2000}.

The plan of the present work is the following. In Section~\ref{sec:diffeq} we obtain a differential equation for a wave function involving only dimensionless quantities; in Subsection~\ref{subsec:schreq} we focus on the one-dimensional case and apply point canonical transformations in order to attain a Schrödinger equation for a new wave function in a space where the mass is constant. In Section~\ref{sec:massprof} we analyze, assuming the absence of external potential, the following five mass profiles: the  {soliton-like} and reciprocal biquadratic profiles (Subsection~\ref{subsec:massprof1}), the reciprocal quadratic profile (Subsection~\ref{subsec:massprof2}), the exponential profile (Subsection~\ref{subsec:massprof3}) and the parabolic profile (Subsection~\ref{subsec:massprof4}). In each case we obtain the effective potentials and the full spectrum of energy and exact analytical solutions. We also plot the probability densities associated with them.  {In Section~\ref{sec:pmdh} we solve the problem of a one-dimensional double heterostructure using a parabolic PDM particle inside the heterojunction among substances characterized by different mass carriers.} Finally, in Section~\ref{sec:conc} we present our conclusions.

%%%%%
%% Section 2 The differential equation for the wave function
%%%%%
 
\section{The dynamical equation for the wave function}
\label{sec:diffeq}

We now rewrite the kinetic Hamiltonian above in order to make explicit the ambiguity sector,
\begin{subequations}
    \label{eq:2-1}
    \begin{equation}
        \label{eq:2-1a}
        \hat{T}_{ {a,b}}(\boldsymbol{r})=\frac{1}{2M}\hat{\boldsymbol{p}}^2-\frac{1}{2M}\left(\frac{1}{M}\hat{\boldsymbol{p}}M\right)\cdot\hat{\boldsymbol{p}}+U_{a,b}(\boldsymbol{r})\;,
    \end{equation}
    where
    \begin{equation}
        \label{eq:2-1b}
        U_{a,b}(\boldsymbol{r})\equiv\frac{1}{2M}\left[\frac{a+b}{2}\hat{\boldsymbol{p}}\left(\frac{1}{M}\hat{\boldsymbol{p}}M\right)-\left(ab+\frac{a+b}{2}\right){\left(\frac{1}{M}\hat{\boldsymbol{p}}M\right)}^2\right]
    \end{equation}
\end{subequations}
behaves like a potential term of kinetic origin (it is noteworthy that, among the Hamiltonians here analyzed, the BDD ordering is the only one for which the ambiguity term is zero).

Should we add an external potential \(V(\boldsymbol{r})\) to \(\hat{T}_{ {a,b}}(\boldsymbol{r})\) we would obtain the following differential equation for the wave function \(\Psi(\boldsymbol{r})\) of the PDM particle 
\begin{multline}
    \label{eq:2-2}
	-\frac{\hbar^2}{2M}{\boldsymbol{\nabla}}^2\Psi(\boldsymbol{r})+\frac{\hbar^2}{2M}\frac{\boldsymbol{\nabla}M}{M}\cdot\boldsymbol{\nabla}\Psi(\boldsymbol{r})+\left(U_{a,b}(\boldsymbol{r})+V(\boldsymbol{r})\right)\Psi(\boldsymbol{r})=E\Psi(\boldsymbol{r})\;,\\
	U_{a,b}(\boldsymbol{r})=-\frac{\hbar^2}{2M}\left[\frac{a+b}{2}\boldsymbol{\nabla}\cdot\left(\frac{\boldsymbol{\nabla}M}{M}\right)-\left(ab+\frac{a+b}{2}\right){\left(\frac{\boldsymbol{\nabla}M}{M}\right)}^2\right]\;.
\end{multline}
This is a generalized Schrödinger equation and it is a very different differential equation from the ordinary constant-mass one.

We will transform the above equation by \(\boldsymbol{r}\rightarrow\epsilon\boldsymbol{r}\), \(\Psi(\boldsymbol{r})\rightarrow\epsilon^{-D/2}\psi(\boldsymbol{r})\) and \(M(\boldsymbol{r})\rightarrow m_0m(\boldsymbol{r})\), where \(\boldsymbol{r}\) is, now, dimensionless; $D$ is the number of spatial dimensions; \(\epsilon\) and $m_0$ are positive constants with dimensions of length and mass, respectively; \(\psi(\boldsymbol{r})\) and \(m(\boldsymbol{r})\) are, respectively, the dimensionless and rescaled wave function and mass. The generalized Schrödinger equation becomes
\begin{subequations}
    \label{eq:2-3}
    \begin{equation}
        \label{eq:2-3a}
	    -\frac{1}{m}{\boldsymbol{\nabla}}^2\psi(\boldsymbol{r})+\frac{1}{m}\frac{\boldsymbol{\nabla}m}{m}\cdot\boldsymbol{\nabla}\psi(\boldsymbol{r})+\left(\tilde{U}_{a,b}(\boldsymbol{r})+\tilde{V}(\boldsymbol{r})\right)\psi(\boldsymbol{r})=\tilde{E}\psi(\boldsymbol{r})\;,
    \end{equation}
    where we write simply \(m(\boldsymbol{r})\equiv m\) and
    \begin{multline}
        \label{eq:2-3b}
	    \tilde{E}\equiv\frac{2\epsilon^2m_0}{\hbar^2}E\quad,\quad\tilde{V}(\boldsymbol{r})\equiv\frac{2\epsilon^2m_0}{\hbar^2}V(\epsilon\boldsymbol{r})\quad\text{and}\\
	    \tilde{U}_{a,b}(\boldsymbol{r})\equiv\frac{2\epsilon^2m_0}{\hbar^2}U_{a,b}(\epsilon\boldsymbol{r})=-\frac{1}{m}\left[\frac{a+b}{2}\boldsymbol{\nabla}\cdot\left(\frac{\boldsymbol{\nabla}m}{m}\right)-\left(ab+\frac{a+b}{2}\right){\left(\frac{\boldsymbol{\nabla}m}{m}\right)}^2\right]
    \end{multline}
\end{subequations}
are the dimensionless and conveniently rescaled energy, external potential and ambiguity kinetic potential, respectively.

%%%%%
%% Subsection 2.1 A new space for a new Schrödinger equation
%%%%%

\subsection{ {The Schrödinger equation for a new wave function in modified space-coordinates}}
\label{subsec:schreq}

For the one-dimensional case, when \(\hat{p}\equiv-i\hbar {\diff{}{x}}\) and \(m\equiv m(x)\), we will perform the change of variable
\begin{subequations}
\label{eq:2.1-1}
	\begin{equation}
	\label{eq:2.1-1a}
		 {\diff{z}{x}}=m^{1/2}\Leftrightarrow z(x)=\int^x\sqrt{m(\mathscr{x})}d\mathscr{x}\;,
	\end{equation}
	 mapping \(\RR\ni x\mapsto z\in\mathcal{D}\subset\RR\), and define a function \(\zeta(z)\) such that
	\begin{equation}
	\label{eq:2.1-1b}
	        \psi(x)=m^{1/4}\ \zeta\!\left(z(x)\right)\;.
	\end{equation}
\end{subequations}
Using this in eq.~\eqref{eq:2-3a} we obtain a Schrödinger equation for a particle with constant mass,
\begin{subequations}
\label{eq:2.1-2}
	\begin{equation}
	\label{eq:2.1-2a}
		-\zeta''(z)+V_{a,b}(z)\zeta(z)=\tilde{E}\zeta(z)\;,
	\end{equation}
	 with a dimensionless effective potential given by
	\begin{equation}
	\label{eq:2.1-2b}
		V_{a,b}(z)\equiv\tilde{V}\left(x(z)\right)+\frac{1}{m}\left[-\frac{1}{2}\left(a+b+\frac{1}{2}\right){\left(\frac{m'}{m}\right)}'+\left(ab+\frac{a+b}{2}+\frac{3}{16}\right){\left(\frac{m'}{m}\right)}^2\right]\;,
	\end{equation}
\end{subequations}
where, for simplicity, we write \(m'\equiv\left. {\diff{m}{x}}\right|_{x(z)}\) and \(m\equiv m\left(x(z)\right)\). Note that for the MM ordering, the effective potential stays simply \(V_{a,b}(z)=\tilde{V}\!\left(x(z)\right)\).

Since here we focus on the absence of external potentials, we will take \(\tilde{V}(x)=0\) everywhere. When the effective potentials $V_{a,b}(z)$ are symmetric around some point, we can write the general solution of eq.~\eqref{eq:2.1-2a}, \(\zeta_{a,b}(z)\), as a linear combination (LC) of particular symmetric and antisymmetric solutions around the same point. Once obtained the solutions that obey the boundary conditions, we use eqs.~\eqref{eq:2.1-1} to write the eigenfunctions in $x$ space (our final goal).

 {It can be seen that the normalization condition for \(\Psi(x)\) remains also valid for \(\psi(x)\)}. On the other hand, by eqs.~\eqref{eq:2.1-1} it is easy to see that \(\sqabs{\psi(x)}dx=\sqabs{\zeta(z)}dz\), which allows the interpretation of \(\zeta(z)\) as a Schrödinger wave function in $z$ space, normalized according to \(\int_{\mathcal{D}}\sqabs{\zeta(z)}dz=1\), where \(\sqabs{\zeta(z)}\) is a probability density.

%For some mass profiles, \emph{e.g.}, \(m(x)={\left(\frac{\alpha+x^2}{1+x^2}\right)}^2\), you cannot get an analytic solution by PCT. For these alternative % cases other techniques must be used \cite{plastino:etal:1999,roy:roy:2002,ycruz:negro:nieto:2008}.

%%%%%
%% Section 3 The mass profiles
%%%%%

\section{The mass profiles}
\label{sec:massprof}

In what follows, we present a variety of non-uniform continuous mass profiles that are found in the literature, some of which we cite along this paper. We study the differential equations resulting from considering such masses for the five most relevant Hamiltonian orderings. First we compute the effective potential in each case, and then we proceed to their analytical treatment. In every case we will obtain the exact full energy spectrum and the corresponding eigensolutions. We analyze the boundary conditions and display the probability densities of a representative number of states. 

%%%%%
%% Subsection 3.1 The  {soliton-like} and reciprocal biquadratic profiles
%%%%%

\subsection{The  {soliton-like} and reciprocal  {biquadratic} profiles}
\label{subsec:massprof1}

Let us focus on two specific mass functions: the  {soliton-like} profile\footnote{  {This profile can be particularly appropriate when hyperbolic potentials are considered. Also, it would possibly be the natural choice in the nonlinear Schrodinger equation with PDM where soliton solutions are present}.} \cite{cunha:christiansen:2013,christiansen:cunha:2013,christiansen:cunha:2014},
\begin{subequations}
    \label{eq:3.1-1}
    \begin{equation}
        \label{eq:3.1-1a}
        M(x)=m_0\sech^2\left(\frac{x}{\epsilon}\right)\Leftrightarrow m(x)=\sech^2x\;,
    \end{equation}
    and the reciprocal  {biquadratic} profile,
    \begin{equation}
        \label{eq:3.1-1b}
        M(x)=m_0\left[1+{\left(\frac{x}{\epsilon}\right)}^2\right]^{-2}\Leftrightarrow m(x)=\left(1+x^2\right)^{-2}\;.
    \end{equation}
\end{subequations}
As will become clear in what follows it will be convenient to treat these two different profiles together.

%%%%%
%% Subsubsection 3.1.1 The effective potential
%%%%%

\subsubsection{The effective potential}

For the so-called  {soliton-like} and reciprocal  {biquadratic} mass distributions, eq.~\eqref{eq:2.1-1a} results in
\begin{equation}
\label{eq:3.1-2}
	\begin{cases}
		\sech x=\cos z &\text{(for the  {soliton-like} profile)}\\
		x=\tan z &\text{(for the reciprocal biquadratic profile)}
	\end{cases}\;,
\end{equation}
both mapping \(\RR\ni x\mapsto z\in\left(-\frac{\pi}{2},\frac{\pi}{2}\right)\). In this case, the effective potentials  {--} eq.~\eqref{eq:2.1-2b}  {--} 
 {parameterized} with subindexes $a$ and $b$ are both given by
\begin{subequations}
    \label{eq:3.1-3}
    \begin{equation}
        \label{eq:3.1-3a}
        V_{a,b}(z)=\omega_{a,b}\tan^2z+V_{a,b}^{(0)}\;,
    \end{equation}
    where
    \begin{equation}
        \label{eq:3.1-3b}
        \omega_{a,b}\equiv4ab+2(a+b)+\frac{3}{4}\quad\text{and}\quad V_{a,b}^{(0)}\equiv V_{a,b}(0)=a+b+\frac{1}{2}
    \end{equation}
    for the  {soliton-like} mass and
    \begin{equation}
        \label{eq:3.1-3c}
        \omega_{a,b}\equiv16ab+6(a+b)+2\quad\text{and}\quad V_{a,b}^{(0)}\equiv V_{a,b}(0)=2(a+b)+1
    \end{equation}
\end{subequations}
for the reciprocal biquadratic mass (see Table~\ref{tab:potef1}).

\begin{longtable}{|>{\centering}m{0.15\textwidth}|>{\centering}m{0.15\textwidth}|>{\centering}m{0.15\textwidth}|>{\centering}m{0.15\textwidth}|m{0.15\textwidth}<{\centering}|}
	\caption{Values of \(\omega_{a,b}\) and $V_{a,b}^{(0)}$ for the most relevant orderings and for each mass profile.}
	\label{tab:potef1} \\
	\hline
						                    &   \multicolumn{2}{c|}{\textbf{ {soliton-like} mass}} &   \multicolumn{2}{c|}{\textbf{Reciprocal biquadratic mass}} \\
	\cline{2-5}
	\multirow[c]{-2}{*}{\textbf{Orderings}}	&   {\boldmath\(\omega_{a,b}\)}	&   {\boldmath$V_{a,b}^{(0)}$}	        &   {\boldmath\(\omega_{a,b}\)}	& {\boldmath$V_{a,b}^{(0)}$} \\
	\hline
	BDD					                    &   $3/4$                       &   $1/2$                               &   $2$                         & $1$ \\
	\hline
	GW					                    &   $-5/4$                      &   $-1/2$                              &   $-4$                        & $-1$ \\
	\hline
	ZK					                    &   $-1/4$                      &   $-1/2$                              &   $0$                         & $-1$ \\
	\hline
	LK					                    &   $-1/4$                      &   $0$                                 &   $-1$                        & $0$ \\
	\hline
	MM					                    &   $0$                         &   $0$                                 &   $0$                         & $0$ \\
	\hline
\end{longtable}

\begin{figure}[H]
	\includegraphics[width=0.5\textwidth]{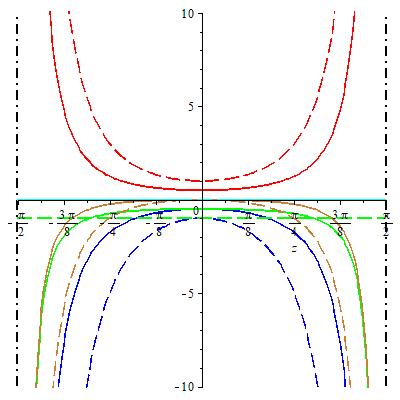}
	\caption{Shape of $V_{a,b}(z)$ for the  {soliton-like} (solid lines) and reciprocal biquadratic (dashed lines) profiles and for the BDD, GW, ZK, LK and MM orderings (in red, blue, green, golden and cyan, respectively).}
	\label{fig:1}
\end{figure}

The parameter \(\omega_{a,b}\) determines the global behaviour of $V_{a,b}(z)$: when \(\omega_{a,b}>0\), $V_{a,b}(z)$ is an infinite potential well; when \(\omega_{a,b}<0\), $V_{a,b}(z)$ is a bottomless barrier potential; finally, when \(\omega_{a,b}=0\), $V_{a,b}(z)$ is constant. The graphics of $V_{a,b}(z)$ can be seen in Fig.~\ref{fig:1}.

%%%%%
%% Subsubsection 3.1.2 Solutions
%%%%%

\subsubsection{Solutions}

Now we define the variable $y$ by means of
\begin{subequations}
\label{eq:3.1-4}
	\begin{equation}
	\label{eq:3.1-4a}
		y=\sin^2z\;,
	\end{equation}
which maps \(\left(-\frac{\pi}{2},\frac{\pi}{2}\right)\ni z\mapsto y\in\left[0,1\right)\), and define a function \(\xi(y)\) by
	\begin{equation}
	\label{eq:3.1-4b}
	        \zeta\!\left(z(y)\right)=y^\mu\left(1-y\right)^\nu\xi(y)\quad,\quad\mu,\nu\in\CC\;.
    \end{equation}
\end{subequations}

Using this in eq.~\eqref{eq:2.1-2a} we obtain the differential equation
\begin{multline}
    \label{eq:3.1-5}
	y\left(1-y\right)\xi''(y)+\left[2\mu+\frac{1}{2}-\left(2\mu+2\nu+1\right)y\right]\xi'(y)+\\
	+\left[\frac{\mu^2-\frac{\mu}{2}}{y}+\frac{\nu^2-\frac{\nu}{2}-\frac{\omega_{a,b}}{4}}{1-y}+\frac{\mathcal{E}_{a,b}+\omega_{a,b}}{4}-\left(\mu+\nu\right)^2\right]\xi(y)=0\;,
\end{multline}
where \(\mathcal{E}_{a,b}\equiv\tilde{E}-V_{a,b}^{(0)}\). \(\mathcal{E}_{a,b}\) thus corresponds to the energy levels measured with respect to the value of the effective potential at the origin. 

Choosing \(\mu\) and \(\nu\) such that\footnote{It can be seen that both values of \(\nu_{a,b}\) lead to the same solution in $z$ space; we will use the negative sign for convenience in the analysis of the boundary conditions.}
\begin{equation}
    \label{eq:3.1-6}
    \mu^2-\frac{\mu}{2}=0 \Rightarrow \mu_\pm=\frac{1\pm1}{4} \quad \text{and} \quad \nu^2-\frac{\nu}{2}-\frac{\omega_{a,b}}{4}=0 \Rightarrow \nu_{a,b}=\frac{1}{4}\left(1-\sqrt{1+4\omega_{a,b}}\right)\;,
\end{equation}
the above differential equation becomes
\begin{multline}
    \label{eq:3.1-7}
	y\left(1-y\right)\xi''(y)+\left[2\mu_\pm+\frac{1}{2}-\left(2\mu_\pm+2\nu_{a,b}+1\right)y\right]\xi'(y)-\\
	-\left[\left(\mu_\pm+\nu_{a,b}\right)^2-\frac{\mathcal{E}_{a,b}+\omega_{a,b}}{4}\right]\xi(y)=0\;.
\end{multline}
This is a hypergeometric Gauss equation,
\begin{equation}
    \label{eq:3.1-8}
    \left\{y\left(1-y\right) {\ddiff{}{y}}+\left[\gamma-\left(\alpha+\beta+1\right)y\right] {\diff{}{y}}-\alpha\beta\right\}\hypergeom{\alpha,\beta;\gamma;y}=0\;,
\end{equation}
where \(\hypergeom{\alpha,\beta;\gamma;y}\) is the hypergeometric Gauss function with parameters
\begin{equation}
    \label{eq:3.1-9}
    \gamma_\pm\equiv2\mu_\pm+\frac{1}{2}=1\pm\dfrac{1}{2} \quad , \quad \alpha_{a,b}^\pm\equiv\mu_\pm+\nu_{a,b}+\frac{1}{2}\sqrt{\mathcal{E}_{a,b}+\omega_{a,b}} \quad\text{and}\quad \beta_{a,b}^\pm\equiv\mu_\pm+\nu_{a,b}-\frac{1}{2}\sqrt{\mathcal{E}_{a,b}+\omega_{a,b}}\;.
\end{equation}

 {For each value of \(\mu\), one of the solutions around $y=0$ is \(\xi_{a,b}^\pm(y)=\hypergeom{\alpha_{a,b}^\pm,\beta_{a,b}^\pm;\gamma_\pm;y}\). Since \(\gamma_\pm\notin\ZZ\), a second independent solution is \(\bar{\xi}_{a,b}^\pm(y)=y^{1-\gamma_\pm}\hypergeom{\alpha_{a,b}^\pm+1-\gamma_\pm,\beta_{a,b}^\pm+1-\gamma_\pm;2-\gamma_\pm;y}\).} By means of eqs.~\eqref{eq:3.1-4} we obtain the two independent solutions (even and odd) in $z$ space:
\begin{equation}
\label{eq:3.1-10}
	\zeta_{a,b}^\pm(z)={\left(\sin z\right)}^{2\mu_\pm}{\left(\cos z\right)}^{2\nu_{a,b}}\hypergeom{\alpha_{a,b}^\pm,\beta_{a,b}^\pm;\gamma_\pm;\sin^2z}\;.
\end{equation}
The solutions in $x$ space are obtained using eqs.~\eqref{eq:2.1-1b}, \eqref{eq:3.1-2} and \eqref{eq:3.1-10}. For the  {soliton-like} mass, one has
\begin{subequations}
\label{eq:3.1-11}
	\begin{align}
	\label{eq:3.1-11a}
		\psi_{a,b}^\pm(x)	&={\left(\tanh x\right)}^{2\mu_\pm}{\left(\sech x\right)}^{2\nu_{a,b}+\frac{1}{2}}\hypergeom{\alpha_{a,b}^\pm,\beta_{a,b}^\pm;\gamma_\pm;\tanh^2x}\nonumber\\
					&={\left(\tanh x\right)}^{2\mu_\pm}{\left(\sech x\right)}^{2\nu_{a,b}+\frac{1}{2}}\times\nonumber\\
					&\times\hypergeom{\mu_\pm+\nu_{a,b}+\frac{1}{2}\sqrt{\mathcal{E}_{a,b}+\omega_{a,b}},\mu_\pm+\nu_{a,b}-\frac{1}{2}\sqrt{\mathcal{E}_{a,b}+\omega_{a,b}};1\pm\frac{1}{2};\tanh^2x}\;,
	\end{align}
	 and for the reciprocal biquadratic mass, 
	\begin{align}
	\label{eq:3.1-11b}
		\psi_{a,b}^\pm(x)	&=x^{2\mu_\pm}{\left(1+x^2\right)}^{-\left(\mu_\pm+\nu_{a,b}+\frac{1}{2}\right)}\hypergeom{\alpha_{a,b}^\pm,\beta_{a,b}^\pm;\gamma_\pm;\frac{x^2}{1+x^2}}\nonumber\\
					&=x^{2\mu_\pm}{\left(1+x^2\right)}^{-\left(\mu_\pm+\nu_{a,b}+\frac{1}{2}\right)}\times\nonumber\\
					&\times\hypergeom{\mu_\pm+\nu_{a,b}+\frac{1}{2}\sqrt{\mathcal{E}_{a,b}+\omega_{a,b}},\mu_\pm+\nu_{a,b}-\frac{1}{2}\sqrt{\mathcal{E}_{a,b}+\omega_{a,b}};1\pm\frac{1}{2};\frac{x^2}{1+x^2}}\;.
	\end{align}
\end{subequations}

Eqs.~\eqref{eq:3.1-11} show that if we compare two orderings having the same value of \(\omega_{a,b}\) (as the ZK and LK orderings with  {soliton-like} mass, or the ZK and MM orderings with reciprocal biquadratic mass) the wave functions distinguish themselves only by \(\mathcal{E}_{a,b}\). Therefore, in this case, if we choose energy levels such that \(\tilde{E}-V_{a,b}^{(0)}\) is the same for the two Hamiltonians, the solutions for both orderings will be also identical.

%%%%%
%% Subsubsection 3.1.3 The energy spectra
%%%%%

\subsubsection{The energy spectra}

The $z$ space solutions must obey the boundary conditions \(\zeta_{a,b}\left(\pm\frac{\pi}{2}\right)=0\).  {We distinguish} three cases depending on the sign of \(\omega_{a,b}\). As resulting from the squeeze theorem we have  {\(\displaystyle \lim_{|z|\rightarrow{\frac{\pi}{2}}^-}{\left(\cos z\right)}^{2\nu_{a,b}}=0\) for \(\omega_{a,b}<0\), $1$ for \(\omega_{a,b}=0\) and \(\infty\) for \(\omega_{a,b}>0\)}, 
%\[\lim_{|z|\rightarrow{\frac{\pi}{2}}^-}{\left(\cos z\right)}^{2\nu_{a,b}}=
%\begin{cases}
%	0 & \text{for} \ \ \ \  \omega_{a,b}<0\\
%	1 & \text{for} \ \ \ \  \omega_{a,b}=0\\
%	\infty & \text{for} \ \ \ \ \omega_{a,b}>0
%\end{cases}\;,\]
and thus, for \(\omega_{a,b}\geq0\) we must impose \(\hypergeom{\alpha_{a,b}^\pm,\beta_{a,b}^\pm;\gamma_\pm;1}=0\). This is the case of MM and BDD orderings for the  {soliton-like} mass and of BDD, ZK  {and} MM orderings for the reciprocal biquadratic mass. 
When \(\Re{\left(\gamma_\pm-\alpha_{a,b}^\pm-\beta_{a,b}^\pm\right)}=\Re{\left(\frac{1}{2}\sqrt{1+4\omega_{a,b}}\right)}>0\), we can use Gauss summation theorem \cite{bailey:1973} to rewrite
\begin{equation}
    \label{eq:3.1-12}
    \hypergeom{\alpha_{a,b}^\pm,\beta_{a,b}^\pm;\gamma_\pm;1}=\frac{\Gamma\!\left(1\pm\frac{1}{2}\right)\Gamma\!\left(\frac{1}{2}\sqrt{1+4\omega_{a,b}}\right)}{\Gamma\!\left(\gamma_\pm-\alpha_{a,b}^\pm\right)\Gamma\!\left(\gamma_\pm-\beta_{a,b}^\pm\right)}\;.
\end{equation}
As a consequence, the boundary condition implies \(\gamma_\pm-\alpha_{a,b}^\pm=-\mathscr{k}\), \(\mathscr{k}\in\ZZ_+\) (\(\mathscr{k}=0,1,2,\ldots\)) which, in turn, results in a discrete non-degenerate energy spectrum for the general solution. Such spectrum reads
\begin{equation}
\label{eq:3.1-13}
	\tilde{E}_{a,b;\mathscr{n}}=\mathscr{n}^2+2\left(1-2\nu_{a,b}\right)\mathscr{n}+1-2\nu_{a,b}+V_{a,b}^{(0)} \quad , \quad \mathscr{n}\in\ZZ_+\;,
\end{equation}
where \(\mathscr{n}\equiv2\mathscr{k}\) for the solutions \(\psi_{a,b}^-(x)\) (even solutions) and \(\mathscr{n}\equiv2\mathscr{k}+1\) for the solutions \(\psi_{a,b}^+(x)\) (odd solutions). The ground state is given by \(\tilde{E}_{a,b;0}=1-2\nu_{a,b}+V_{a,b}^{(0)}\) which is greater than $V_{a,b}^{(0)}$ as expected.

On the other hand, for \(\omega_{a,b}<0\), the effective potentials are bottomless barriers and the boundary conditions are automatically respected for any energy level. The spectrum is determined by the derivative \( {{\zeta_{a,b}^\pm}'(z)}\) at the border
\begin{equation}
    \label{eq:3.1-14}
    \lim_{z\rightarrow\pm\frac{\pi}{2}}{\zeta_{a,b}^\pm}'(z)=-2\nu_{a,b}\hypergeom{\alpha_{a,b}^\pm,\beta_{a,b}^\pm;\gamma_\pm;1}\lim_{z\rightarrow\pm\frac{\pi}{2}}{\left(\cos{z}\right)}^{2\nu_{a,b}-1}
\end{equation}
 {which is divergent}. Actually, for \(-\frac{1}{4}<\omega_{a,b}<0\) one can avoid the singularity by cancellation of an identical factor in the hypergeometric function. In this case we would get the same spectrum given by eq.~\eqref{eq:3.1-13}. This is a highly unexpected result for a bottomless barrier seems to behave like an infinite potential well. The same cannot be done for  {\(\omega_{a,b}\leq-\frac{1}{4}\)} because the transcendental equation \(\hypergeom{\alpha_{a,b}^\pm,\beta_{a,b}^\pm;\gamma_\pm;1}=0\) has no solutions for \(\tilde{E}\);  {therefore the GW, ZK and LK orderings with soliton-like mass} and GW and LK orderings with reciprocal biquadratic mass do not have physically acceptable states and must be discarded.

Now we can write the discrete eigenstates (even and odd) compactly in terms of \(\mathscr{k}\):
\begin{subequations}
\label{eq:3.1-15}
	\begin{equation}
	\label{eq:3.1-15a}
		\psi_{a,b;\mathscr{k}}^\pm(x)={\left(\tanh x\right)}^{2\mu_\pm}{\left(\sech x\right)}^{2\nu_{a,b}+\frac{1}{2}}\hypergeom{2\mu_\pm+\frac{1}{2}+\mathscr{k},2\nu_{a,b}-\frac{1}{2}-\mathscr{k};1\pm\frac{1}{2};\tanh^2x}
	\end{equation}
	for the  {soliton-like} mass, and
	\begin{equation}
	\label{eq:3.1-15b}
		\psi_{a,b;\mathscr{k}}^\pm(x)=x^{2\mu_\pm}{\left(1+x^2\right)}^{-\left(\mu_\pm+\nu_{a,b}+\frac{1}{2}\right)}\hypergeom{2\mu_\pm+\frac{1}{2}+\mathscr{k},2\nu_{a,b}-\frac{1}{2}-\mathscr{k};1\pm\frac{1}{2};\frac{x^2}{1+x^2}}
	\end{equation}
\end{subequations}
for the reciprocal biquadratic mass.

%%%%%
%% Subsubsection 3.1.4 Graphics of the probability densities
%%%%%

\subsubsection{Graphics of the probability densities}

\begin{figure}[H]
	\subfigure[Plots of \(\rho_{-1/4,-1/4;\mathscr{k}}^\pm(x)\) for \(\mathscr{k}=0\), $1$ and $2$ (solid, dashed and dotted lines, respectively). \label{fig:2a}]{\includegraphics[width=0.45\textwidth]{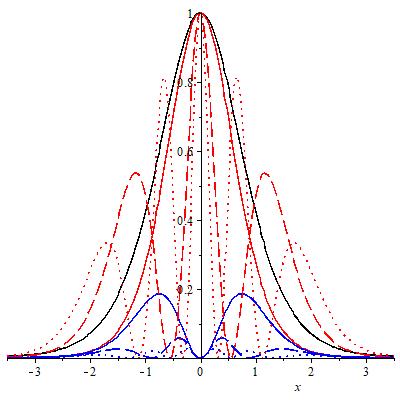}} \ \ 
	\subfigure[Plots of \(\rho_{0,0;\mathscr{k}}^\pm(x)\) for \(\mathscr{k}=0\), $1$ and $2$ (solid, dashed and dotted lines, respectively). \label{fig:2b}]{\includegraphics[width=0.45\textwidth]{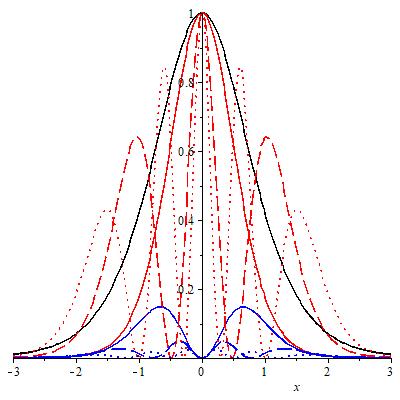}}
	\caption{Graphics of probability densities associated with the even (in red) and odd (in blue) unnormalized solutions for the  {soliton-like} mass and for the \subref{fig:2a} MM and \subref{fig:2b} BDD orderings. The solid black line corresponds to \(m(x)=\sech^2x\).}
	\label{fig:2}
\end{figure}

\begin{figure}[H]
	\subfigure[Plots of \(\rho_{-1/2,-1/2;\mathscr{k}}^\pm(x)\) and \(\rho_{-1/4,-1/4;\mathscr{k}}^\pm(x)\) for \(\mathscr{k}=0\), $1$ and $2$ (solid, dashed and dotted lines, respectively). \label{fig:3a}]{\includegraphics[width=0.45\textwidth]{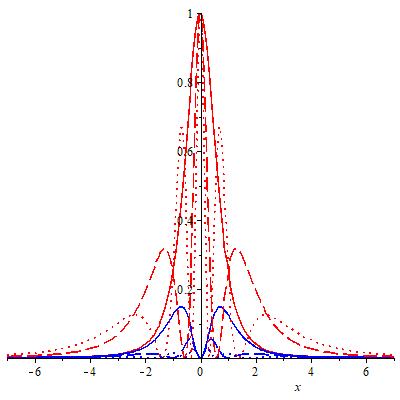}} \ \
	\subfigure[Plots of \(\rho_{0,0;\mathscr{k}}^\pm(x)\) for \(\mathscr{k}=0\), $1$ and $2$ (solid, dashed and dotted lines, respectively). \label{fig:3b}]{\includegraphics[width=0.45\textwidth]{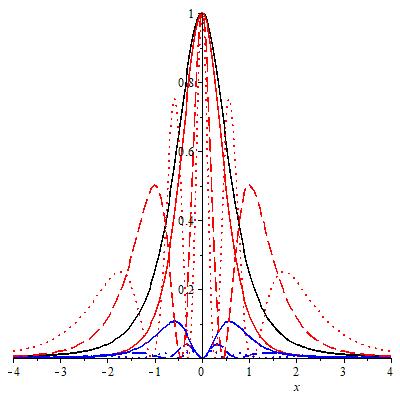}}
	\caption{Graphics of probability densities associated with the even (in red) and odd (in blue) unnormalized solutions for the reciprocal biquadratic mass and for the \subref{fig:3a} ZK and MM and \subref{fig:3b} BDD orderings. The solid black line corresponds to \(m(x)={\left(1+x^2\right)}^{-2}\).}
	\label{fig:3}
\end{figure}

The probability densities associated with each particular solution, \(\rho_{a,b}^\pm(x)\equiv\sqabs{\psi_{a,b}^\pm(x)}\) and \(\rho_{a,b;\mathscr{k}}^\pm(x)\equiv\sqabs{\psi_{a,b;\mathscr{k}}^\pm(x)}\), are plotted in Figs.~\ref{fig:2} and \ref{fig:3} (for the  {soliton-like} and reciprocal biquadratic masses, respectively).

From Fig.~\ref{fig:3a} we can see that \(\rho_{-1/2,-1/2;0}^-(x)=\rho_{-1/4,-1/4;0}^-(x)=m(x)\), which is easily verified algebraically using the identity \(\hypergeom{\alpha,\beta;\alpha;y}=\left(1-y\right)^{-\beta}\).

%%%%%
%% Subsection 3.2 The reciprocal quadratic profile
%%%%%

\subsection{The reciprocal quadratic profile}
\label{subsec:massprof2}

Now we study the reciprocal quadratic mass \cite{mathews:lakshmanan:1974,mathews:lakshmanan:1975,yu:dong:2004,ycruz:negro:nieto:2007,karthiga:etal:2017,dacosta:gomez:portesi:2020},
\begin{equation}
    \label{eq:3.2-1}
    M(x)=m_0{\left[1+{\left(\frac{x}{\epsilon}\right)}^2\right]}^{-1}\Leftrightarrow m(x)={\left(1+x^2\right)}^{-1}\;.
\end{equation}

%%%%%
%% Subsubsection 3.2.1 The effective potential
%%%%%

\subsubsection{The effective potential}

For this mass, eq.~\eqref{eq:2.1-1a} results in
\begin{equation}
\label{eq:3.2-2}
	x=\sinh z\;,
\end{equation}
 mapping \(\RR\ni x\mapsto z\in\RR\). The effective potential  {--} eq.~\eqref{eq:2.1-2b}  {--} is
 \begin{subequations}
    \label{eq:3.2-3}
    \begin{equation}
        \label{eq:3.2-3a}
        V_{a,b}(z)=\omega_{a,b}\sech^2z+V_{a,b}^{(\infty)}=V_{a,b}^{(0)}-\omega_{a,b}\tanh^2z\;,
    \end{equation}
    where we define (see Table~\ref{tab:potef2})
    \begin{equation}
        \label{eq:3.2-3b}
        \omega_{a,b}\equiv\frac{1}{4}-4ab \quad,\quad V_{a,b}^{(0)}\equiv V_{a,b}(0)=a+b+\frac{1}{2} \quad\text{and}\quad V_{a,b}^{(\infty)}\equiv V_{a,b}\left(\pm\infty\right)=4ab+a+b+\frac{1}{4}\;.
    \end{equation}
\end{subequations}
\begin{table}[H]
    \centering
    \begin{tabular}{|>{\centering}m{0.2\textwidth}|>{\centering}m{0.2\textwidth}|>{\centering}m{0.2\textwidth}|m{0.2\textwidth}<{\centering}|}
        \hline
        \textbf{Orderings}	&   {\boldmath\(\omega_{a,b}\)}	&   {\boldmath$V_{a,b}^{(0)}$}	&   {\boldmath\(V_{a,b}^{(\infty)}\)} \\
        \hline
        BDD			        &   $1/4$                       &   $1/2$                       &   $1/4$ \\
	    \hline
		GW			        &   $1/4$                       &   $-1/2$                      &   $-3/4$ \\
    	\hline
		ZK			        &   $-3/4$                      &   $-1/2$                      &   $1/4$ \\
    	\hline
		LK			        &   $1/4$                       &   $0$                         &   $-1/4$ \\
    	\hline
		MM			        &   $0$                         &   $0$                         &   $0$ \\
    	\hline
    \end{tabular}
    \caption{Values of \(\omega_{a,b}\), $V_{a,b}^{(0)}$ and \(V_{a,b}^{(\infty)}\) for each ordering, for the reciprocal quadratic mass.}
    \label{tab:potef2}
\end{table}

We observe that $V_{a,b}(z)$ is a finite potential well when \(\omega_{a,b}<0\) and a finite barrier when \(\omega_{a,b}>0\); when \(\omega_{a,b}=0\), $V_{a,b}(z)$ is just uniform. Furthermore, since \(V_{a,b}^{(0)}-V_{a,b}^{(\infty)}=\omega_{a,b}\), we can interpret \(\left|\omega_{a,b}\right|\) as the amplitude of the potential well or barrier.

\begin{figure}[H]
	\includegraphics[width=0.5\textwidth]{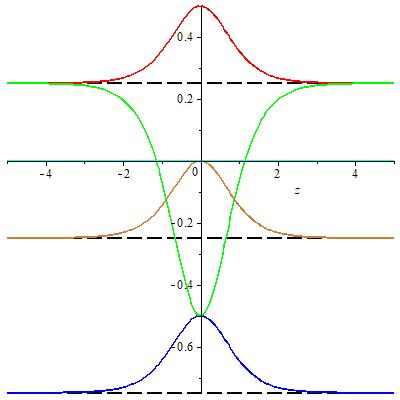}
	\caption{Graphics of $V_{a,b}(z)$ for the reciprocal quadratic profile and for the BDD, GW, ZK, LK and MM orderings (in red, blue, green, golden and cyan, respectively).}
	\label{fig:4}
\end{figure}

The graphics of $V_{a,b}(z)$ can be seen in Fig.~\ref{fig:4}.

%%%%%
%% Subsubsection 3.2.2 Solutions
%%%%%

\subsubsection{Solutions}

We change variables again by defining
\begin{subequations}
\label{eq:3.2-4}
	\begin{equation}
	\label{eq:3.2-4a}
		y=\tanh^2z\;,
	\end{equation}
 mapping \(\RR\ni z\mapsto y\in\left[0,1\right)\), and define a function \(\xi(y)\) by
	\begin{equation}
	\label{eq:3.2-4b}
		\zeta\!\left(z(y)\right)=y^\mu\left(1-y\right)^\nu\xi(y) \quad,\quad \mu,\nu\in\CC\;.
	\end{equation}
\end{subequations}

Thus, eq.~\eqref{eq:2.1-2a} becomes the following differential equation:
\begin{multline}
    \label{eq:3.2-5}
	y\left(1-y\right)\xi''(y)+\left[2\mu+\frac{1}{2}-\left(2\mu+2\nu+\frac{3}{2}\right)y\right]\xi'(y)+\\
	+\left[\frac{\mu^2-\frac{\mu}{2}}{y}+\frac{\nu^2+\frac{\mathcal{E}_{a,b}}{4}}{1-y}-\frac{\omega_{a,b}}{4}-\left(\mu+\nu\right)^2-\frac{\mu+\nu}{2}\right]\xi(y)=0\;,
\end{multline}
where we defined \(\mathcal{E}_{a,b}\equiv \tilde{E}-V_{a,b}^{(\infty)}\). Thus \(\mathcal{E}_{a,b}\) represents the energy levels measured with respect to value of the potential at infinite. In this case we choose \(\mu\) and \(\nu\) such that
\begin{equation}
    \label{eq:3.2-6}
    \mu^2-\frac{\mu}{2}=0\Rightarrow\mu_\pm=\frac{1\pm1}{4} \quad\text{and}\quad \nu^2+\frac{\mathcal{E}_{a,b}}{4}=0\Rightarrow\nu_{a,b}=-\frac{1}{2}\sqrt{-\mathcal{E}_{a,b}}\;,
\end{equation}
the above equation becomes
\begin{multline}
    \label{eq:3.2-7}
	y\left(1-y\right)\xi''(y)+\left[2\mu_\pm+\frac{1}{2}-\left(2\mu_\pm+2\nu_{a,b}+\frac{3}{2}\right)y\right]\xi'(y)-\\
	-\left[{\left(\mu_\pm+\nu_{a,b}\right)}^2+\frac{\mu_\pm+\nu_{a,b}}{2}+\frac{\omega_{a,b}}{4}\right]\xi(y)=0\;,
\end{multline}
which is a Gauss hypergeometric equation with parameters
\begin{equation}
    \label{eq:3.2-8}
    \gamma_\pm\equiv2\mu_\pm+\dfrac{1}{2}=1\pm\dfrac{1}{2} \quad,\quad \alpha_{a,b}^\pm\equiv\mu_\pm+\nu_{a,b}+\frac{1}{4}+\sqrt{ab} \quad\text{and}\quad \beta_{a,b}^\pm\equiv\mu_\pm+\nu_{a,b}+\frac{1}{4}-\sqrt{ab}\;.
\end{equation}

Since \(\gamma_\pm\notin\ZZ\), then for both values of \(\mu\) the previous equation has the same two independent solutions of Subsection~\ref{subsec:massprof1} around $y=0$, \(\xi_{a,b}^\pm(y)\) and \(\bar{\xi}_{a,b}^\pm(y)\). Combined with eqs.~\eqref{eq:3.2-4}, these yield the two independent solutions in $z$ space,
\begin{equation}
\label{eq:3.2-9}
	\zeta_{a,b}^\pm(z)={\left(\tanh z\right)}^{2\mu_\pm}{\left(\sech z\right)}^{2\nu_{a,b}}\hypergeom{\alpha_{a,b}^\pm,\beta_{a,b}^\pm;\gamma_\pm;\tanh^2z}\;.
\end{equation}
The solutions in $x$ space are obtained by combining the eqs.~\eqref{eq:2.1-1b}, \eqref{eq:3.2-2} and \eqref{eq:3.2-9}:
\begin{align}
\label{eq:3.2-10}
	\psi_{a,b}^\pm(x)	&=x^{2\mu_\pm}{\left(1+x^2\right)}^{-\left(\mu_\pm+\nu_{a,b}+\frac{1}{4}\right)}\hypergeom{\alpha_{a,b}^\pm,\beta_{a,b}^\pm;\gamma_\pm;\frac{x^2}{1+x^2}}\nonumber\\
				&=x^{2\mu_\pm}{\left(1+x^2\right)}^{-\left(\mu_\pm+\nu_{a,b}+\frac{1}{4}\right)}\times\nonumber\\
				&\times\hypergeom{\mu_\pm+\nu_{a,b}+\frac{1}{4}+\sqrt{ab},\mu_\pm+\nu_{a,b}+\frac{1}{4}-\sqrt{ab};1\pm\frac{1}{2};\frac{x^2}{1+x^2}}\;.
\end{align}

Once again, when we compare two orderings for which \(\omega_{a,b}\) are equal (the case of BDD, GW and LK orderings), the solutions are distinguished by the value of \(\mathcal{E}_{a,b}\). Therefore, if for such orderings we measure the energy levels with respect to \(V_{a,b}^{(\infty)}\), the solutions are identical. Interestingly, for \(\nu_{a,b}=\frac{1}{2}\sqrt{-\mathcal{E}_{a,b}}\), we would be led to equivalent results.

%%%%%
%% Subsubsection 3.2.3 The energy spectra
%%%%%

\subsubsection{The energy spectra}

 {In this case} the normalization condition for \(\zeta_{a,b}(z)\) demands \(\zeta_{a,b}^\pm\left(\pm\infty\right)<\infty\). Since  {\(\displaystyle \lim_{z\rightarrow\pm\infty}{\left(\sech z\right)}^{2\nu_{a,b}}\) is undefined for \(\mathcal{E}_{a,b}>0\), equal to $1$ for \(\mathcal{E}_{a,b}=0\) and infinity for \(\mathcal{E}_{a,b}<0\),} we can separate our analysis in the following three cases:
\begin{enumerate}
	\item \(\mathcal{E}_{a,b}>0\Leftrightarrow\tilde{E}>V_{a,b}^{(\infty)}\) (scattering states): \(\zeta_{a,b}(z)\) oscillates indefinitely, modulated by \(\hypergeom{\alpha_{a,b}^\pm,\beta_{a,b}^\pm;\gamma_\pm;\tanh^2z}\), but \(\psi_{a,b}(x)\) is normalizable because
	\begin{equation}
	    \label{eq:3.2-11}
	    \lim_{x\rightarrow\pm\infty}{m^{1/4}}=\lim_{x\rightarrow\pm\infty}{\left(1+x^2\right)^{-1/4}}=0\;.
	\end{equation}
	The energy spectrum is continuous;
	\item \(\mathcal{E}_{a,b}=0\Leftrightarrow\tilde{E}=V_{a,b}^{(\infty)}\) (trivial state): \(\zeta_{a,b}(z)\) reduces to the trivial solution and is thus irrelevant; 
	\item \(\mathcal{E}_{a,b}<0\Leftrightarrow\tilde{E}<V_{a,b}^{(\infty)}\) (bound states): in this case we must impose that \(\hypergeom{\alpha_{a,b}^\pm,\beta_{a,b}^\pm;\gamma_\pm;1}=0\) and, therefore, since \(\Re{\left(\gamma_\pm-\alpha_{a,b}^\pm-\beta_{a,b}^\pm\right)}=\Re{\left(\sqrt{-\mathcal{E}_{a,b}}\right)}>0\), we can use again the Gauss summation theorem and take \(\gamma_\pm-\alpha_{a,b}^\pm=-\mathscr{k}\), \(\mathscr{k}\in\ZZ_+\) (\(\mathscr{k}=0,1,2,\ldots\)). With this we obtain, for the general solution, a discrete non-degenerate energy spectrum
	\begin{subequations}
	\label{eq:3.2-12}
		\begin{equation}
		\label{eq:3.2-12a}
			\tilde{E}_{a,b;\mathscr{n}}=-\mathscr{n}^2+\left(4\sqrt{ab}-1\right)\mathscr{n}+2\sqrt{ab}-\frac{1}{2}+V_{a,b}^{(0)}\;,
		\end{equation}
		where \(\mathscr{n}\equiv2\mathscr{k}\) for the solutions \(\psi_{a,b}^-(x)\) (even solutions) and \(\mathscr{n}\equiv2\mathscr{k}+1\) for the solutions \(\psi_{a,b}^+(x)\) (odd solutions). We also verify that these states are restricted to the interval
		\begin{equation}
		\label{eq:3.2-12b}
			0\leq\mathscr{n}<2\sqrt{ab}-\frac{1}{2} \Leftrightarrow 0\leq\mathscr{k}<\sqrt{ab}-\mu_\pm-\frac{1}{4}\;.
		\end{equation}
	\end{subequations}
\end{enumerate}

Put in terms of \(\omega_{a,b}\):
\begin{itemize}
\item when \(\omega_{a,b}\geq0\), solutions are all of scattering type;
	\item when \(-2\leq\omega_{a,b}<0\), there is only one  {bound state (the ground state), all the others are of scattering type}; 
	\item when \(\omega_{a,b}<-2\),  {both bound and scattering states exist, and} the general solution is given by the LC of both particular solutions.
\end{itemize}

 {Synthesizing}, when the effective potential is a finite barrier (\(\omega_{a,b}>0\)) or uniform (\(\omega_{a,b}=0\)), there are no bound states. On the other hand, when the effective potential is a finite well (\(\omega_{a,b}<0\)), the bound states are discretized, with energy levels given by eqs.~\eqref{eq:3.2-12}. Written in terms of \(\mathscr{k}\), eq.~\eqref{eq:3.2-10} reads
\begin{multline}
\label{eq:3.2-13}
	\psi_{a,b;\mathscr{k}}^\pm(x)=x^{2\mu_\pm}{\left(1+x^2\right)}^{-\left(\mathscr{k}+2\mu_\pm+\frac{1}{2}-\sqrt{ab}\right)}\times\\
	\times\hypergeom{\mathscr{k}+2\mu_\pm+\frac{1}{2},\mathscr{k}+2\mu_\pm+\frac{1}{2}-2\sqrt{ab};1\pm\frac{1}{2};\frac{x^2}{1+x^2}}\;.
\end{multline}
In this case, when we compare two orderings that have the same \(\omega_{a,b}\), the solutions can be distinguished 
by the values of \(\mathscr{k}\).

%%%%%
%% Subsubsection 3.2.4 Graphics of the probability densities
%%%%%

\subsubsection{Graphics of the probability densities}

\begin{figure}[H]
	\subfigure[Graphics of \(\rho_{-1/2,-1/2;0}^-(x)\) (solid red line) and of \(\rho_{-1/2,-1/2}^\pm(x)\) for \(\tilde{E}=1/2\) (dashed lines).\label{fig:5a}]{\includegraphics[width=0.32\textwidth]{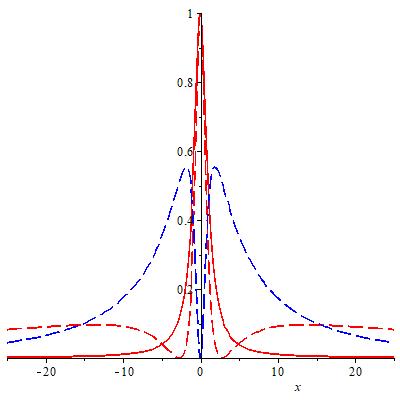}} \ \ 
	\subfigure[Graphics of \(\rho_{-1/4,-1/4}^\pm(x)\) for \(\tilde{E}=1/32\) and $1$ (solid and dashed lines, respectively).\label{fig:5b}]{\includegraphics[width=0.32\textwidth]{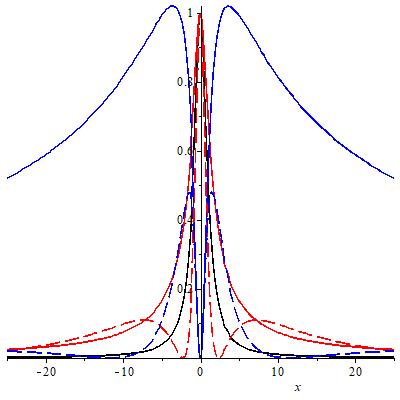}} \ \ 
	\subfigure[Graphics of \(\rho_{0,0}^\pm(x)\), \(\rho_{-1,0}^\pm(x)\) and \(\rho_{0,-1/2}^\pm(x)\) for \(\tilde{E}=V_{a,b}^{(\infty)}+1/32\) and \(V_{a,b}^{(\infty)}+1\) (solid and dashed lines, respectively).\label{fig:5c}] {\includegraphics[width=0.32\textwidth]{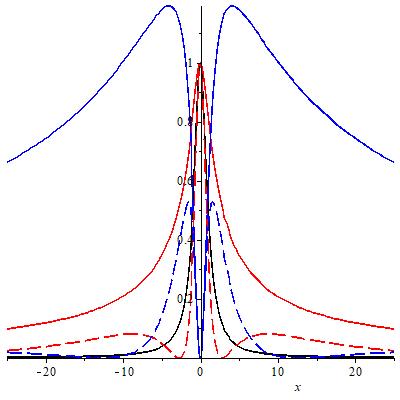}}
	\caption{Graphics of the probability densities associated to even (in red) and odd (in blue) unnormalized solutions for the reciprocal quadratic mass and for the \subref{fig:5a} ZK, \subref{fig:5b} MM and \subref{fig:5c} BDD, GW and LK orderings. The solid black line is for the curve \(m(x)=\left(1+x^2\right)^{-1}\).}
	\label{fig:5}
\end{figure}

For this mass profile, the ZK ordering is the only one having bound states; more specifically, just the ground state\footnote{See Table~\ref{tab:potef2} and compare with eqs.~\eqref{eq:3.2-12}} (the graphics of both \(\rho_{-1/2,-1/2;0}^-(x)\) and \(\rho_{-1/2,-1/2}^\pm(x)\) for \(\tilde{E}=1/2\) are plotted in Fig.~\ref{fig:5a}). For the other orderings the graphics of \(\rho_{a,b}^\pm(x)\) are plotted in Figs.~\ref{fig:5b}  {and} \ref{fig:5c}; in this case the BDD, ZK and LK orderings fit the same figure.

In Fig.~\ref{fig:5a} we verify that \(\rho_{-1/2,-1/2;0}^-(x)=m(x)\), which can be verified using the identity \(\hypergeom{\alpha,\beta;\alpha;y}=\left(1-y\right)^{-\beta}\).

%%%%%
%% Subsection 3.3 The exponential profile
%%%%%

\subsection{The exponential profile}
\label{subsec:massprof3}

Let us take the following exponential profile \cite{desouzadutra:almeida:2000,gonul:etal:2002,midya:roy:2012}:
\begin{equation}
    \label{eq:3.3-1}
    M(x)=m_0e^{-2|x|/\epsilon}\Leftrightarrow m(x)=e^{-2|x|}\;,
\end{equation}
which vanishes asymptotically in both directions.

%%%%%
%% Subsubsection 3.3.1 The effective potential
%%%%%

\subsubsection{The effective potential}

In this case  {eq.~\eqref{eq:2.1-1a}} implies the relationship
\begin{equation}
\label{eq:3.3-2}
    e^{-|x|}=1-|z|\;,
\end{equation}
mapping \(\RR\ni x\mapsto z\in\left(-1,1\right)\). The integration constants for each branch were chosen in such a way that \(\RR_-^*\ni x\mapsto z\in\left(-1,0\right)\) and \(\RR_+\ni x\mapsto z\in\left[-1,1\right)\). The effective potential  {--} eq.~\eqref{eq:2.1-2b}  {--} is
\begin{equation}
    \label{eq:3.3-3}
    V_{a,b}(z)=\frac{\omega_{a,b}}{{\left(1-|z|\right)}^2} \quad , \quad \omega_{a,b}\equiv4ab+2\left(a+b\right)+\frac{3}{4},
\end{equation}
which is continuous throughout the domain (interestingly, \(\omega_{a,b}\) is the same one found for the  {soliton-like} profile). When \(\omega_{a,b}\neq0\) it has a critical point \(V_{a,b}(0)=\omega_{a,b}\) and two singularities, at \(z=\pm1\), which depend on the signal of \(\omega_{a,b}\): if \(\omega_{a,b}>0\) then $V_{a,b}(z)$ is an infinite well; if \(\omega_{a,b}<0\) it is bottomless barrier.

It is easy to see that when different orderings have the same value of \(\omega_{a,b}\) (as is the case with the ZK and LK orderings) they also share the same effective potential and therefore we expect the same solutions for the same energies.

\begin{figure}[H]
    \includegraphics[width=0.5\textwidth]{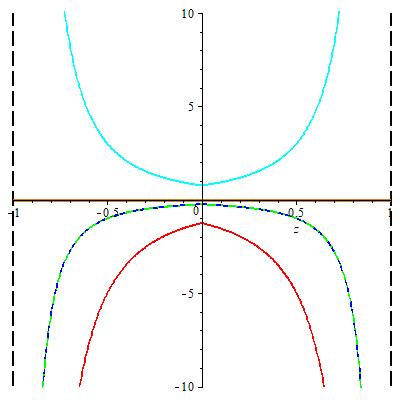}
    \caption{Graphics of $V_{a,b}(z)$ referring to the exponential mass for the GW, ZK, LK, MM and BDD orderings (in red, blue, green, golden and cyan, respectively). Note the overlap between $V_{-1/2,-1/2}(z)$ and $V_{0,-1/2}(z)$}.
    \label{fig:6}
\end{figure}

The graphics of $V_{a,b}(z)$ can be seen in Fig.~\ref{fig:6}.

%%%%%
%% Subsubsection 3.3.2 Solutions
%%%%%

\subsubsection{Solutions}

This time we change the variables by
\begin{subequations}
\label{eq:3.3-4}
	\begin{equation}
	\label{eq:3.3-4a}
        y=\nu\left(1-|z|\right) \quad , \quad \nu\in\CC^*\;,
	\end{equation}
	 mapping \(\left(-1,1\right)\ni z\mapsto y\in\left\{\mathscr{y}\in\CC^*,|\mathscr{y}|\leq|\nu|\right\}\) (at the perimeter \(|y|=|\nu|\) ($z=0$), the solutions in $z$ space can be obtained by continuity) and define a function \(\xi(y)\) such that
	\begin{equation}
	\label{eq:3.3-4b}
    	\zeta\!\left(z(y)\right)=\left(\frac{y}{\nu}\right)^{1/2}\xi(y)\;.
	\end{equation}
\end{subequations}

With this, eq.~\eqref{eq:2.1-2a} transforms into\footnote{Here, \(\II_+^*\) denotes the set of elements written in the form $ih$, where \(h\in\RR_+^*\).}
\begin{subequations}
    \label{eq:3.3-5}
    \begin{equation}
        \label{eq:3.3-5a}
        y^2\xi''(y)+y\xi'(y)+ \left(\frac{\tilde{E}}{\nu^2}y^2-\alpha_{a,b}^2\right)\xi(y)=0 \quad , \quad \alpha_{a,b}\equiv\frac{1}{2}\sqrt{1+4\omega_{a,b}}\in\RR_+\cup\II_+^*\;.
    \end{equation}
    Choosing \(\nu=\sqrt{\tilde{E}}\) for \(\tilde{E}\neq0\), and \(\nu=1\) for \(\tilde{E}=0\), the above equation becomes
    \begin{equation}
        \label{eq:3.3-5b}
        \begin{cases}
            y^2\xi''(y)+y\xi'(y)+\left(y^2-\alpha_{a,b}^2\right)\xi(y)=0 & \tilde{E}\neq0 \\
            y^2\xi''(y)+y\xi'(y)-\alpha_{a,b}^2\xi(y)=0 & \tilde{E}=0
        \end{cases}\;,
    \end{equation}
\end{subequations}
the first line is equivalent to a regular Bessel’s equation of order \(\alpha_{a,b}\) and the second to a 2nd order homogeneous Euler-Cauchy equation. % \cite[pp. 166 -- 172]{zill:2017}.

The two independent solutions of each differential equation are
\begin{equation}
    \label{eq:3.3-6}
    \xi_{a,b}^{(1)}(y)=
    \begin{cases}
        J_{\alpha_{a,b}}(y) & \tilde{E}\neq0 \\
        y^{\alpha_{a,b}} & \tilde{E}=0
    \end{cases} \quad \text{and} \quad \xi_{a,b}^{(2)}(y)=
    \begin{cases}
        H_{\alpha_{a,b}}^{(1)}(y) & \tilde{E}\neq0 \\
        \begin{cases}
            y^{-\alpha_{a,b}} & \alpha_{a,b}\neq0 \\
            \ln{y} & \alpha_{a,b}=0 \\
        \end{cases} & \tilde{E}=0
    \end{cases}\;,
\end{equation}
where \(J_\alpha(y)\) and \(H_\alpha^{(1)}(y)\) are, respectively, the 1st kind Bessel and Hankel functions of order \(\alpha\). Combining them with the eqs.~\eqref{eq:3.3-4} we obtain the solutions in $z$ space:
\begin{subequations}
\label{eq:3.3-7}
    \begin{equation}
    \label{eq:3.3-7a}
        \zeta_{a,b}^{(1)}(z)=
        \begin{cases}
            \sqrt{1-|z|}J_{\alpha_{a,b}}\!\left(\sqrt{\tilde{E}}\left(1-|z|\right)\right) & \tilde{E}\neq0 \\
            {\left(1-|z|\right)}^{\frac{1}{2}+{\alpha_{a,b}}} & \tilde{E}=0
        \end{cases}
    \end{equation}
    and
    \begin{equation}
    \label{eq:3.3-7b}
        \zeta_{a,b}^{(2)}(z)=
        \begin{cases}
            \sqrt{1-|z|}H_{\alpha_{a,b}}^{(1)}\!\left(\sqrt{\tilde{E}}\left(1-|z|\right)\right) & \tilde{E}\neq0 \\
            \begin{cases}
                {\left(1-|z|\right)}^{\frac{1}{2}-{\alpha_{a,b}}} & {\alpha_{a,b}}\neq0 \\
                \sqrt{1-|z|}\ln{\left(1-|z|\right)} & {\alpha_{a,b}}=0
            \end{cases} & \tilde{E}=0
        \end{cases}\;.
    \end{equation}
\end{subequations}

%%%%%
%% Subsubsection 3.3.3 The energy spectra
%%%%%

\subsubsection{The energy spectra}

The general solution in $z$ space, \(\zeta_{a,b}(z)\), must obey the conditions \(\zeta_{a,b}\left(\pm1\right)=0\) and \(\zeta_{a,b}'(0)=0\) (the second one is obtained from continuity of wave function derivative; note in eqs.~\eqref{eq:3.3-7} that functions are even).

Taking into account the asymptotic behavior of Bessel functions for small arguments and with the aid of identity
\begin{equation}
    \label{eq:3.3-8}
    \mathcal{C}_\alpha'(y)=\mathcal{C}_{\alpha-1}(y)-\frac{\alpha}{y}\mathcal{C}_\alpha(y)\;,
\end{equation}
where \(\mathcal{C}_\alpha(y)\) denotes the functions \(J_\alpha(y)\) or \(H_\alpha^{(1)}(y)\), we conclude that when \(\alpha_{a,b}\in\II_+^* \Leftrightarrow \omega_{a,b}<-\frac{1}{4}\) there is no solution meeting the above conditions. On the other hand, when \(\alpha_{a,b}\geq0 \Leftrightarrow \omega_{a,b}\geq-\frac{1}{4}\) only eq.~\eqref{eq:3.3-7a} is a physically valid solution. It presents a discrete energy spectrum \(\tilde{E}_{a,b;\mathscr{n}}\equiv k_{a,b;\mathscr{n}}^2\), where \(k_{a,b;\mathscr{n}}>0\) is the \(\mathscr{n}\)th positive solution of the transcendental equation
\begin{equation}
\label{eq:3.3-9}
    2k_{a,b}J_{\alpha_{a,b}-1}\!\left(k_{a,b}\right)+\left(1-2\alpha_{a,b}\right)J_{\alpha_{a,b}}\!\left(k_{a,b}\right)=0\;,
\end{equation}
Since the exchange \(k_{a,b;\mathscr{n}}\rightarrow-k_{a,b;\mathscr{n}}\) produces the same energy spectrum, we find that the eigenstates show double degeneracy.

In the $x$ space the solutions are obtained by combining the eqs.~\eqref{eq:2.1-1b}, \eqref{eq:3.3-2} and \eqref{eq:3.3-7}, considering eq.~\eqref{eq:3.3-9}
\begin{equation}
\label{eq:3.3-10}
    \psi_{a,b;\mathscr{n}}(x)=e^{-|x|}J_{\alpha_{a,b}}\!\left(k_{a,b;\mathscr{n}}e^{-|x|}\right) \quad , \quad \mathscr{n}\in\ZZ_+^* \quad , \quad \omega_{a,b}\geq-\frac{1}{4}\;.
\end{equation}
Looking at eq.~\eqref{eq:3.3-10} we see that eigenstates can be distinguished from each other by the ordering, represented by \(\alpha_{a,b}\), as well as the energy level \(\tilde{E}_{a,b;\mathscr{n}}\equiv k_{a,b;\mathscr{n}}^2\). Thus, the eigenstates with same energy are indistinguishable between two orderings with the same \(\omega_{a,b}\), as expected.

The value of \(k_{a,b;\mathscr{n}}=\sqrt{\tilde{E}_{a,b;\mathscr{n}}}\) can be calculated numerically for each ordering with \(\omega_{a,b}\geq-\frac{1}{4}\) from eq.~\eqref{eq:3.3-9} (see Table \ref{tab:enerexp}).

\vskip 2cm
\begin{longtable}{|c|c|c|c|}
	\caption{Three first values of \(k_{a,b;\mathscr{n}}\) for the BDD, ZK/LK and MM orderings. The GW ordering has no valid solutions.}
	\label{tab:enerexp} \\ 
	\hline
						& \multicolumn{3}{c|}{\boldmath\(k_{a,b;\mathscr{n}}\)} \\
	\cline{2-4}
	\multirow[c]{-2}{*}{\textbf{Orderings}}	& {\boldmath\(\mathscr{n}=1\)}	& {\boldmath\(\mathscr{n}=2\)}	& {\boldmath\(\mathscr{n}=3\)} \\
	\hline
	ZK/LK				                    & $0.9407705639$                &   $3.959371185$				& $7.086380848$ \\
	\hline
	MM					                    & $1.570796327$                 &   $4.712388980$			    & $7.853981634$ \\
	\hline
	BDD					                    & $2.165871271$                 &   $5.427433202$			    & $8.595426306$ \\
	\hline
\end{longtable}

It is worth noting that a system with exponential mass has only bound states, even for orderings with \(-\frac{1}{4}\leq\omega_{a,b}<0\), corresponding to potentials $V_{a,b}(z)$ which are bottomless barriers. This contradicts the common sense that there should be only scattering states and a continuous and inferiorly unlimited spectrum. As we can see, in this case a bottomless barrier behaves like infinite potential well.

The GW ordering is automatically removed for having no valid solution.

%%%%%
%% Subsubsection 3.3.4 Graphics of the probability densities
%%%%%

\subsubsection{Graphics of the probability densities}

\begin{figure}[H]
	\subfigure[\label{fig:7a}]{\includegraphics[width=0.32\textwidth]{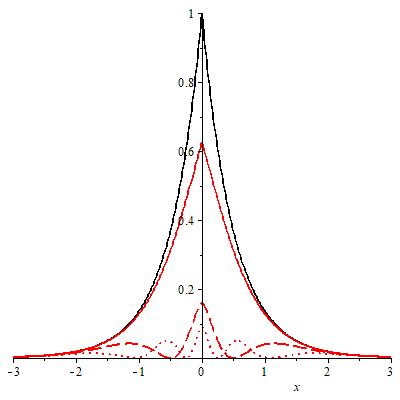}} \
	\subfigure[\label{fig:7b}]{\includegraphics[width=0.32\textwidth]{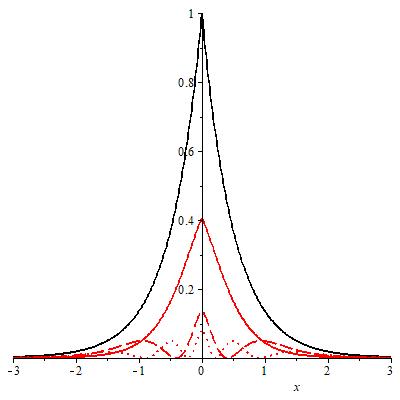}} \
	\subfigure[\label{fig:7c}]{\includegraphics[width=0.32\textwidth]{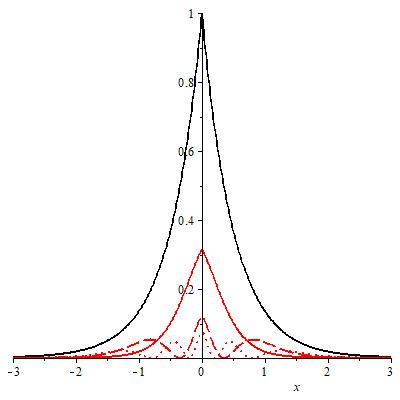}}
	\caption{Graphics of the probability densities associated with eq.~\eqref{eq:3.3-10} (unnormalized solutions) for the exponential mass and for the \subref{fig:7a} ZK and LK, \subref{fig:7b} MM and \subref{fig:7c} BDD orderings, with \(\mathscr{n}=1\), $2$ and $3$ (solid, dashed and dotted lines, respectively). The solid black line corresponds to $m(x)=e^{-2|x|}$.}
	\label{fig:7}
\end{figure}

The graphics of \(\rho_{a,b;\mathscr{n}}(x)\equiv\sqabs{\psi_{a,b;\mathscr{n}}(x)}\) are potted in Fig.~\ref{fig:7}. The probability densities associated with ZK and LK orderings are plotted in a same picture because they are identical.

Fig.~\ref{fig:7} shows that although solutions in $z$ space have continuous derivatives, this is not true for solutions in $x$ space.  In fact, taking the derivative of eq.~\eqref{eq:3.3-10} and considering eq.~\eqref{eq:3.3-9} we conclude that
\begin{equation}
    \label{eq:3.3-11}
    \lim_{x\rightarrow0^\pm}\psi_{a,b;\mathscr{n}}'(x)=\pm\frac{1}{k_{a,b;\mathscr{n}}}\left(\frac{1}{2}-k_{a,b;\mathscr{n}}\right)J_{\alpha_{a,b}}\!\left(k_{a,b;\mathscr{n}}\right)\;,
\end{equation}
\emph{i.e.}, the derivatives of the solutions are continuous at the origin only if \(k_{a,b;\mathscr{n}}=\frac{1}{2}\) or if \(k_{a,b;\mathscr{n}}\) is some zero of the 1st type Bessel function, which is not necessarily true.

%%%%%
%% Subsection 3.4 The parabolic profile
%%%%%

\subsection{The parabolic profile}
\label{subsec:massprof4}

Let us focus on the analysis of a parabolic mass density \cite{nascimento:guedes:2014,schmidt:2006,yu:dong:2004},
\begin{equation}
    \label{eq:3.4-1}
    M(x)=m_0{\left(\frac{2x}{\epsilon}\right)}^2\Leftrightarrow m(x)=4x^2\;.
\end{equation}
Since it is an indefinitely growing profile, we restrict it to a finite region \(|x|<\ell,\;\ell>0\) which could possibly represent a heterostructure. This leaves the solutions free of boundary conditions.

%%%%%
%% Subsubsection 3.4.1 The effective potential
%%%%%

\subsubsection{The effective potential}

For this mass, eq.~\eqref{eq:2.1-1a} results in the transformation
\begin{equation}
\label{eq:3.4-2}
    x^2=|z|\;,
\end{equation}
mapping \(\left(-\ell,\ell\right)\ni x\mapsto z\in\left(-\ell^2,\ell^2\right)\). Eq.~\eqref{eq:2.1-1b} gives the effective potential
\begin{equation}
    \label{eq:3.4-3}
    V_{a,b}(z)=\frac{\omega_{a,b}}{z^2} \quad , \quad \omega_{a,b}\equiv ab+\frac{3}{4}\left(a+b\right)+\frac{5}{16}\;.
\end{equation}

\begin{figure}[H]
    \includegraphics[width=0.5\textwidth]{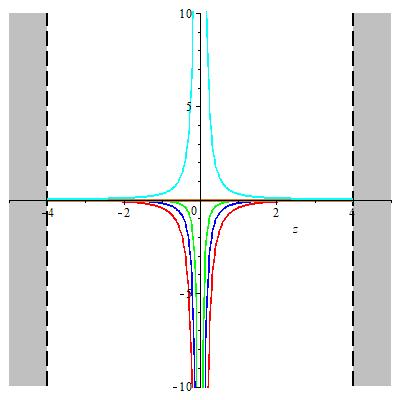}
    \caption{Shape of the effective potentials $V_{a,b}(z)$ for the GW (\(\omega_{-1,0}=-7/16\)), ZK (\(\omega_{-1/2,-1/2}=-3/16\)), LK (\(\omega_{0,-1/2}=-1/16\)), MM (\(\omega_{-1/4,-1/4}=0\)) and BDD (\(\omega_{0,0}=5/16\)) orderings (in red, blue, green, golden and cyan, respectively) and related to a system restricted, in $x$ space, to the interval \(|x|\leq2\).}
    \label{fig:8}
\end{figure}

We observe that this expression is similar to the exponential mass effective potential under transformation \(1-|z|\rightarrow|z|\), (and a different \(\omega_{a,b}\)). Therefore, some analogous considerations can be made. Indeed, depending on the sign of \(\omega_{a,b}\) the potential will behave very differently: if \(\omega_{a,b}>0\) (\(\omega_{a,b}<0\)) then the potential is an infinite barrier (bottomless well) because it diverges to \(+\infty\) (\(-\infty\)) at the origin. The plots of the effective potentials for the different orderings are in Fig.~\ref{fig:8}.

%%%%%
%% Subsubsection 3.4.2 Solutions
%%%%%

\subsubsection{Solutions}

The resemblance in $z$ space of this with the exponential mass system allows the recognition of eqs.~\eqref{eq:3.3-7} as solutions for this case. We just need to make \(1-|z|\rightarrow |z|\) and substitute the definition of \(\omega_{a,b}\) (the \(\alpha_{a,b}\) and \(\omega_{a,b}\) parameters are still related by \(\alpha_{a,b}=\frac{1}{2}\sqrt{1+4\omega_{a,b}}\)).

The analysis of the solutions and their derivatives at the origin, show that there is no solution for \(\alpha_{a,b}\in\II_+^*\cup\left[0,\frac{1}{2}\right)\Leftrightarrow\omega_{a,b}<0\), which means that the orderings whose effective potentials in $z$ space are bottomless wells are automatically excluded for this profile. When \(\alpha_{a,b}=\frac{1}{2}\Leftrightarrow\omega_{a,b}=0\) only the first solution with zero energy is acceptable, and it is reduced to \(\zeta_{a,b}(z)=1\). On the other hand, for \(\alpha_{a,b}>\frac{1}{2}\Leftrightarrow\omega_{a,b}>0\) just the first solution and their derivative converge and are continuous at $z=0$. Thus, for positive \(\omega_{a,b}\) parameters the physically acceptable solutions are
\begin{equation}
\label{eq:3.4-4}
    \zeta_{a,b}(z)=
    \begin{cases}
        \sqrt{|z|}J_{\alpha_{a,b}}\!\left(\sqrt{\tilde{E}}|z|\right) & \tilde{E}\neq0\\
        |z|^{\frac{1}{2}+\alpha_{a,b}} & \tilde{E}=0
    \end{cases} \quad , \quad \alpha_{a,b}>\frac{1}{2}\;.
\end{equation}
which combined with eqs.~\eqref{eq:2.1-1b}, \eqref{eq:3.4-2} result in the  {$x$} space eigenstates
\begin{equation}
\label{eq:3.4-5}
    \psi_{a,b}(x)=
    \begin{cases}
        |x|^{3/2}J_{\alpha_{a,b}}\!\left(\sqrt{\tilde{E}}x^2\right) & \tilde{E}\neq0\\
        |x|^{\frac{3}{2}+2\alpha_{a,b}} & \tilde{E}=0
    \end{cases} \quad , \quad \omega_{a,b}>0\;.
\end{equation}

%%%%%
%% Subsubsection 3.4.3 The energy spectra
%%%%%

\subsubsection{The energy spectra}

%\emph{A priori}, there are no restrictions or conditions on the eigenvalues related to eigenstates of eq.~\eqref{eq:3.4-5}. This is due to the fact we narrow our system to the interval \(x\in(-\ell,\ell)\). 

If we consider a heterostructure model with arbitrary mass and/or external potential in \(|x|>\ell\), we can study how the potentials $V_{a,b}(z)$ behave in the region \(|z|>\ell^2\). Then, we will be able to find energy levels and eventually evaluate the conditions for discretization or degeneracy resulting from \(\displaystyle{\lim_{z\rightarrow\pm\ell^{\pm}}\zeta_{a,b}(z)}\) and \(\displaystyle{\lim_{z\rightarrow\pm\ell^{\pm}}\zeta_{a,b}'(z)}\).

For example, if we assume that in \(|x|>\ell\) the mass is uniform and the external potential is still zero, then $V_{a,b}(z)=0$ in \(|z|>\ell^2\) and solutions will be all free scattering states. If instead the mass has positive concavity out of the interval \((-\ell^2,\ell^2)\) the effective potential will result in a potential well and it will be possible to have bound states with negative energy above its minimum. 

%%%%%
%% Subsubsection 3.4.4 Graphics of the probability densities
%%%%%

\subsubsection{Graphics of the probability densities}

\begin{figure}[H]
	\includegraphics[width=0.4\textwidth]{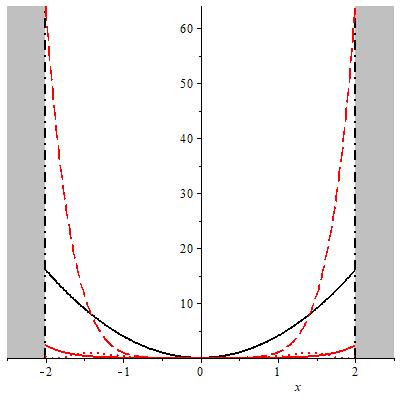}
	\caption{Graphics of BDD \(\rho_{0,0}(x)\) for \(\tilde{E}=-1/32\), $0$ and $1$ (solid, dashed and dotted red line, respectively). The solid black line corresponds to $m(x)=4x^2$. The other orderings are not acceptable.}
	\label{fig:9}
\end{figure}

The plots of the probability densities for the BDD ordering are shown in Fig.~\ref{fig:9}. The previous discussion gives us a plausible reason to illustrate at least one situation with negative eigenenergy. In any case, the displayed sector (\(|x|<\ell\)) shows tunneling states across the potential for parabolic mass distribution and BDD ordering. 

%%%%%
%% Section 4 Parabolic mass in a double heterostructure
%%%%%

\section{ {Double heterostructure: an example}}
\label{sec:pmdh}
 {In order to apply our previous results, we will consider a model of a one-dimensional quantum heterostructure where a charge carrier among two regions has a PDM of parabolic type. The particle mass changes from  $m_0m_1$ to  $m_0m_2$,  $m_{1,2}>0$, along an intermediate region IR of size  \(2\ell\) symmetric about the origin: 
\begin{equation}
    \label{eq:4-1}
    M(x)=
    \begin{cases}
        m_0m_1 & \dfrac{x}{\epsilon}<-\ell \\
        4m_0\eta{\left(\dfrac{x}{\epsilon}+\chi\right)}^2 & \left|\dfrac{x}{\epsilon}\right| \leq \ell \\
        m_0m_2 & \dfrac{x}{\epsilon}>\ell
    \end{cases}
    \Leftrightarrow m=
    \begin{cases}
        m_1 & x<-\ell \\
        4\eta{\left(x+\chi\right)}^2 & \left|x\right| \leq \ell \\
        m_2 & x>\ell
    \end{cases} \;,
\end{equation}
with \(\chi=\frac{\sqrt{m_2}+\sqrt{m_1}}{\sqrt{m_2}-\sqrt{m_1}}\ell\) and \(\eta=\frac{{\left(\sqrt{m_2}-\sqrt{m_1}\right)}^2}{16\ell^2}\).
}

%%%%%
%% Subsection 4.1 Solutions
%%%%%

\subsection{ {Solutions}}
\label{subsec:sols}

 {
For $|x|>\ell$ the mass is constant and the solutions are the well-known ordinary wavefunctions
\begin{subequations}
    \label{eq:4.1-1}
    \begin{equation}
        \label{eq:4.1-1a}
        \psi(x)=
        \begin{cases}
            Ae^{\left|k_1\right|x} & x<-\ell \\
            Be^{-\left|k_2\right|x} & x>\ell
        \end{cases}
      \quad\quad\quad \quad \tilde{E}<0  %\quad \text{(bound states)}
    \end{equation}
    and
    \begin{equation}
        \label{eq:4.1-1b}
        \psi(x)=
        \begin{cases}
            Ae^{ik_1x}+ARe^{-ik_1x} & x<-\ell \\
            ATe^{ik_2x} & x>\ell
        \end{cases}
      \quad \tilde{E}>0% \quad \text{(scattering states)} \;,
    \end{equation}
\end{subequations}
where \(A,B,R,T\in\CC\) ($R$ and $T$ are associated with reflection and transmision coefficients) and \(k_{1,2}\equiv{\left(m_{1,2}\tilde{E}\right)}^{1/2}\).
}  {
In the IR the particular solutions   \(\psi_{a,b}^{(1,2)}(x)\) are those  obtained in the previous section, 
with \(x \rightarrow x+\chi\) and \(\tilde{E} \rightarrow \eta\tilde{E}\). %Note that since the mass never vanishes, 
Now, the boundary conditions allow two independent components and the general solution inside is  \(\psi_{a,b}^\text{in}(x)\) in  \(|x| \leq \ell\),  given by  \(\psi_{a,b}^\text{in}(x)=C^{(1)}\psi_{a,b}^{(1)}(x)+C^{(2)}\psi_{a,b}^{(2)}(x)\), with \(C^{(1,2)}\in\CC\) and   
\begin{subequations}
    \label{eq:4.1-2}
    \begin{equation}
        \label{eq:4.1-2a}
        \psi_{a,b}^{(1)}(x)=
        \begin{cases}
            {|x+\chi|}^{3/2}I_{\alpha_{a,b}}\left({\left(\eta\left|\tilde{E}\right|\right)}^{1/2}{(x+\chi)}^2\right) & \tilde{E}<0 \\
            {|x+\chi|}^{3/2}J_{\alpha_{a,b}}\left({\left(\eta\tilde{E}\right)}^{1/2}{(x+\chi)}^2\right) & \tilde{E}>0
        \end{cases}
    \end{equation}
    and
    \begin{equation}
        \label{eq:4.1-2b}
        \psi_{a,b}^{(2)}(x)=
        \begin{cases}
            {|x+\chi|}^{3/2}K_{\alpha_{a,b}}\left({\left(\eta\left|\tilde{E}\right|\right)}^{1/2}{(x+\chi)}^2\right) & \tilde{E}<0 \\
            {|x+\chi|}^{3/2}H_{\alpha_{a,b}}^{(1)}\left({\left(\eta\tilde{E}\right)}^{1/2}{(x+\chi)}^2\right) & \tilde{E}>0
        \end{cases} \;,
    \end{equation}
\end{subequations}
} {
where we have chosen to write these solutions in terms of modified Bessel functions.}

%%%%%
%% Subsection 4.2 Boundary conditions
%%%%%

\subsection{ {Boundary conditions}}
\label{subsec:boundcond}
 {The boundary conditions are defined by imposing the continuity of 
\(\psi_{a,b}(x)\) and \(\psi_{a,b}'(x)\). We can write these conditions in a compact way as follows. Define the following column vectors
\begin{multline}
    \label{eq:4.2-1}
    \boldsymbol{\vartheta}_{a,b;\tau}\equiv
    \begin{pmatrix}
        \psi_{a,b}^{(1)}\left((-1)^\tau\ell\right) \\
        \psi_{a,b}^{(2)}\left((-1)^\tau\ell\right)
    \end{pmatrix} \quad,\quad
    \dot{\boldsymbol{\vartheta}}_{a,b;\tau}\equiv
    \begin{pmatrix}
        {\psi_{a,b}^{(1)}}'\left((-1)^\tau\ell\right) \\
        {\psi_{a,b}^{(2)}}'\left((-1)^\tau\ell\right)
    \end{pmatrix} \quad,\quad
    \boldsymbol{C}\equiv
    \begin{pmatrix}
        C^{(1)} \\
        C^{(2)}
    \end{pmatrix} \quad\text{and} \\
    \boldsymbol{\kappa}_{a,b;\tau} \equiv \dot{\boldsymbol{\vartheta}}_{a,b;\tau}-(-1)^\tau ik_\tau\boldsymbol{\vartheta}_{a,b;\tau} \,,
\end{multline}
where \(\tau=1\), $2$. 
With \(\gamma^{(1,2)}\in\CC\), we also define
\begin{equation}
    \label{eq:4.2-2}
    {
    \begin{pmatrix}
        \gamma^{(1)} \\
        \gamma^{(2)}
    \end{pmatrix}
    }_\perp\equiv
    \begin{pmatrix}
        \gamma^{(2)} \\
        -\gamma^{(1)}
    \end{pmatrix}
    =
    \begin{pmatrix}
        0   &   1 \\
        -1  &   0
    \end{pmatrix}
    \begin{pmatrix}
        \gamma^{(1)} \\
        \gamma^{(2)}
    \end{pmatrix} \;,
\end{equation}
\emph{i.e.}, index \(\perp\) denotes a rotation of \(\frac{\pi}{2}\) counter clockwise.}

 {
The boundary conditions imply
\begin{subequations}
    \label{eq:4.2-3}
    \begin{equation}
        \label{eq:4.2-3a}
        \begin{cases}
            \begin{pmatrix}
                \boldsymbol{\vartheta}_{a,b;1}^T \\
                \dot{\boldsymbol{\vartheta}}_{a,b;1}^T
            \end{pmatrix}
            \boldsymbol{C}=Ae^{-\ell\left|k_1\right|}
            \begin{pmatrix}
                1 \\
                \left|k_1\right|
            \end{pmatrix} \\
            \begin{pmatrix}
                \boldsymbol{\vartheta}_{a,b;2}^T \\
                \dot{\boldsymbol{\vartheta}}_{a,b;2}^T
            \end{pmatrix}
            \boldsymbol{C}=Be^{-\ell\left|k_2\right|}
            \begin{pmatrix}
                1 \\
                -\left|k_2\right|
            \end{pmatrix}
        \end{cases} \quad \tilde{E}<0%\text{(bound states)} \quad ,
    \end{equation}
    and
    \begin{equation}
        \label{eq:4.2-3b}
        \begin{cases}
            \begin{pmatrix}
                \boldsymbol{\vartheta}_{a,b;1}^T \\
                \dot{\boldsymbol{\vartheta}}_{a,b;1}^T
            \end{pmatrix}
            \boldsymbol{C}=Ae^{-i\ell k_1}
            \begin{pmatrix}
                1 \\
                i k_1
            \end{pmatrix}
            +ARe^{i\ell k_1}
            \begin{pmatrix}
                1 \\
                -i k_1
            \end{pmatrix} \\
            \begin{pmatrix}
                \boldsymbol{\vartheta}_{a,b;2}^T \\
                \dot{\boldsymbol{\vartheta}}_{a,b;2}^T
            \end{pmatrix}
            \boldsymbol{C}=ATe^{i\ell k_2}
            \begin{pmatrix}
                1 \\
                i k_2
            \end{pmatrix}
        \end{cases} \quad \tilde{E}>0%\text{(scattering states)}
    \end{equation}
\end{subequations}
}

 {
(${}^T$ denoting transposition).}
 {
For $\tilde{E}<0$, eqs.~\eqref{eq:4.2-3a} yield $B$ and $C^{(1,2)}$ in terms of $A$,
\begin{subequations}
    \label{eq:4.2-4}
    \begin{equation}
        \label{eq:4.2-4a}
        \boldsymbol{C}_{a,b}=-A_{a,b}e^{-\ell \left|k_1\right|}\frac{\boldsymbol{\kappa}_{a,b;1\perp}}{\dot{\boldsymbol{\vartheta}}_{a,b;1}^T\boldsymbol{\vartheta}_{a,b;1\perp}} \quad \text{and} \quad
        B_{a,b}=-A_{a,b}e^{\ell\left( \left|k_2\right|-\left|k_1\right|\right)}\frac{\boldsymbol{\vartheta}_{a,b;2}^T\boldsymbol{\kappa}_{a,b;1\perp}}{\dot{\boldsymbol{\vartheta}}_{a,b;1}^T\boldsymbol{\vartheta}_{a,b;1\perp}} .
    \end{equation}
$A$ will be determined by normalization, and the discrete energy levels will result from the following transcendental equation
    \begin{equation}
        \label{eq:4.2-4b}
        \boldsymbol{\kappa}_{a,b;1}^T\boldsymbol{\kappa}_{a,b;2\perp}=0  .
    \end{equation}
\end{subequations}
}
 {
For $\tilde{E}>0$ states,  eqs.~\eqref{eq:4.2-3b} yield $C^{(1,2)}$, $R$ and $T$ em terms of the free incident wave amplitude $A$,
\begin{multline}
    \label{eq:4.2-5}
    \boldsymbol{C}_{a,b}=-2Aik_1e^{-i\ell k_1}\frac{\boldsymbol{\kappa}_{a,b;2\perp}}{\boldsymbol{\kappa}_{a,b;1}^T\boldsymbol{\kappa}_{a,b;2\perp}} \quad \text{and} \\
    \begin{pmatrix}
        R_{a,b} \\
        T_{a,b}
    \end{pmatrix}
    =-\frac{e^{-i\ell k_1}}{\boldsymbol{\kappa}_{a,b;1}^T\boldsymbol{\kappa}_{a,b;2\perp}}
    \begin{pmatrix}
        e^{-i\ell k_1}\left(\dot{\boldsymbol{\vartheta}}_{a,b;1}^T-ik_1\boldsymbol{\vartheta}_{a,b;1}^T\right)\boldsymbol{\kappa}_{a,b;2\perp} \\
        2ik_1e^{-i\ell k_2}\dot{\boldsymbol{\vartheta}}_{a,b;2}^T\boldsymbol{\vartheta}_{a,b;2\perp}
    \end{pmatrix}  .
\end{multline}
}

%%%%%
%% Subsection 4.3 Results
%%%%%

\subsection{ {Results}}
\label{subsec:res}
 
 {In order to plot our results, we choose the mass parameters
  \(m_1={m_2}/{2}=0.5\) and \(\ell=1\). We will consider \(m_0\approx m_e\), where $m_e$ is the free electron mass, and \(\epsilon\approx \mu m\) for a typical heterojunction among semiconductors.}
 {In this particular case  there are no bound states since 
there is no solution to eq.~\eqref{eq:4.2-4b}.}  {Instead, we have scattering states and the study of the flux probability results in
}

\begin{figure}[H]
	\subfigure[\label{fig:10a}]{\includegraphics[width=0.32\textwidth]{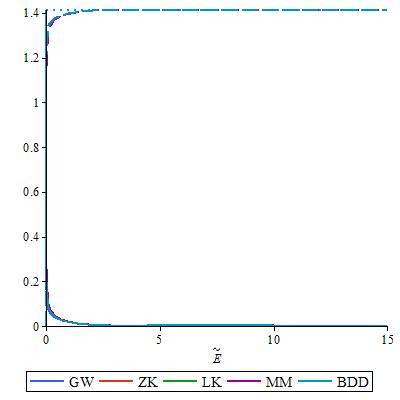}} \
	\subfigure[\label{fig:10b}]{\includegraphics[width=0.32\textwidth]{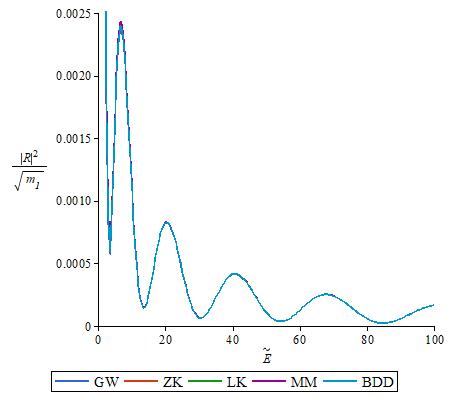}} \
	\subfigure[\label{fig:10c}]{\includegraphics[width=0.32\textwidth]{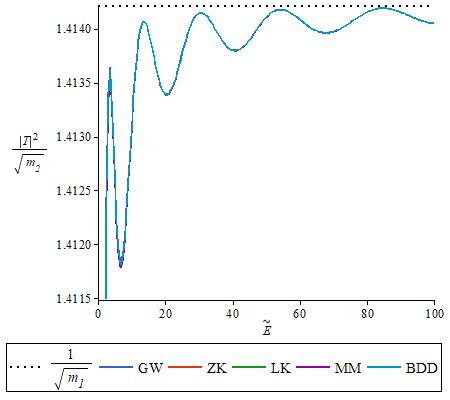}}
	\caption{\subref{fig:10a}  {Plot of \({\left|R_{a,b}\right|}^2/\sqrt{m_1}\) and \({\left|T_{a,b}\right|}^2/\sqrt{m_2}\) (solid and dashed lines, respectively). Pointed line corresponds to  \({\left|R_{a,b}\right|}^2/\sqrt{m_1}+{\left|T_{a,b}\right|}^2/\sqrt{m_2},\) which equals \(1/\sqrt{m_1}\), see eq.~\eqref{eq:4.3-1}. \subref{fig:10b} and \subref{fig:10c} are magnifications of details of Fig.~\ref{fig:10a}.}}
	\label{fig:10}
\end{figure}

 {
\begin{equation}
    \label{eq:4.3-1}
    \frac{{\left|R_{a,b}\right|}^2}{\sqrt{m_1}}+\frac{{\left|T_{a,b}\right|}^2}{\sqrt{m_2}}=\frac{1}{\sqrt{m_1}} \quad  .
\end{equation}
}
 {In Fig.~\ref{fig:10} we can see the behaviour of \({\left|R_{a,b}\right|}^2\) and \({\left|T_{a,b}\right|}^2\) with respect to the energy, and verify the validity of eq.~\eqref{eq:4.3-1}.
}

\begin{figure}[H]
	\subfigure[\label{fig:11a}]{\includegraphics[width=0.49\textwidth]{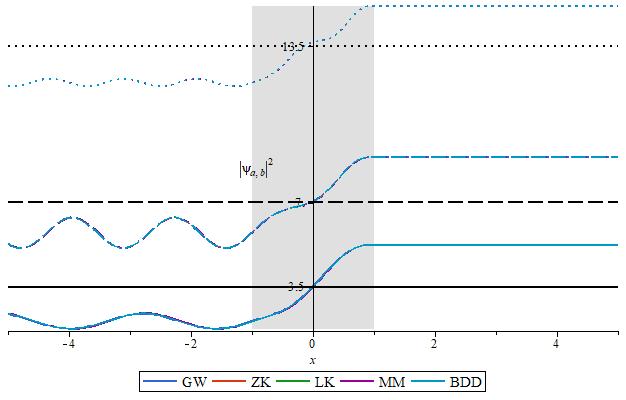}} \
	\subfigure[\label{fig:11b}]{\includegraphics[width=0.49\textwidth]{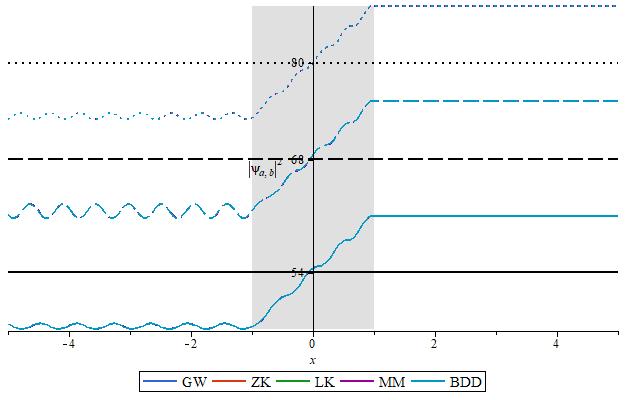}}
	\caption{\subref{fig:11a}  {Plots of \({\left|\psi_{a,b}(x)\right|}^2\) for \(\tilde{E}=3.5\), $7.0$ and $13.5$ (solid, dashed and pointed lines, respectively) and \subref{fig:11b} \(\tilde{E}=54\), $68$ and $80$ (solid, dashed and pointed lines, respectively). The shaded zone is the intermediate region IR where the mass varies with position. The amplitudes are chosen for the sake of visualization and the graphics have been moved up vertically to distinguish energy levels.}.}
	\label{fig:11}
\end{figure}

 {In Fig.~\ref{fig:11} we plot the probability densities for some energy levels .}

%%%%%
%% Section 5 Final discussion
%%%%%

\section{Final discussion}
\label{sec:conc}

According to the canonical commutation rules of quantum mechanics, when a position-dependent mass is considered the quantum particle's Hamiltonian can be built in a variety of forms. This would inevitably result in very different energy spectra and particle eigenstates. In this paper we have focused on the study of a purely kinetic Hamiltonian so as to compare the effects of considering several arrangements of ordering and mass profiles put on an equal footing. For this, we have performed the systematic calculation of the energy eigenvalues and eigenfunctions of Hamiltonians resulting from five different orderings and five distinct spatially depending masses. We have obtained a fully new and large variety of eigenstates arising as a consequence of the different ordering/mass arrangements. Among them, Gauss hypergeometric and Bessel functions related to hyperbolic, exponential or rational functions have been found  {--} see eqs.~\eqref{eq:3.1-11}, \eqref{eq:3.2-10}, \eqref{eq:3.3-9} and \eqref{eq:3.4-5}. In Table~\ref{tab:spectra} we summarize the type of energy spectra of the whole diversity of arrangements. In this table we can see that the reciprocal quadratic mass is the only profile presenting just continuous spectra for every ordering, all with a minimal eigenvalue. For this profile, the only ordering to have an energy gap between the minimum and the continuous band is ZK. The   {soliton-like}, reciprocal biquadratic and exponential profiles admit only discrete spectra and some orderings do not have any allowed physical solutions (GW, ZK and LK for soliton-like, GW and LK for reciprocal biquadratic and only GW for exponential). The parabolic mass is the one profile with less solutions for the whole set of orderings.

Among the different orderings, GW is the only not having physically allowed solutions for any profile but the reciprocal quadratic. MM and BDD, on the other hand, have a full spectrum with a ground state for every mass profile. In the case of discrete spectra, the ZK energy levels are always lower than MM's which in turn are just below those of the BDD ordering {(similar comparisons between energy spectra can be found in \cite{kulikov:2012})}.

%{By comparing the discrete spectra -see eqs.~\eqref{eq:3.1-13}, \eqref{eq:3.2-12} and \eqref{eq:3.3-9} (with Table~\ref{tab:enerexp})- we can classify the corresponding energy levels of the different Hamiltonians for each mass profile; see e.g. \cite{kulikov:2012} for a similar analysis. 

Depending on the arrangement, we have obtained a variety of effective potentials from the pure kinetic PDM Hamiltonian. In Subsections~\ref{subsec:massprof1} and \ref{subsec:massprof3} we have found even bottomless barriers with discrete spectral solutions associated (e.g. ZK and LK orderings with  {soliton-like} mass).   {
Some authors have addressed the subject of bottomless potentials and provided ways of dealing with these results \cite{vachaspati:2002,sous:kawni:2009,demic:etal:2016,ahmed:etal:2018,hu:etal:2018,ahmed:etal:2019,cho:ho:2008:novel,cho:ho:2008:self}. If we assume that free PDM particles mimic constant mass particles in a solid, we can think of these bottomless barriers as those appearing between the infinite potential wells of nuclei or ions in a crystal. This is a very well-known problem of solid-state physics \cite{cohen-tannoudji:diu:laloe:1978}. }
%The continuous spectra could for instance be associated with conduction bands, and the discrete spectra with the lower energy eigenstates of the electrostatic nuclear potential wells. 

We have also verified that some ordering/mass arrangements are discarded for not showing any acceptable solution (neither bound nor scattering eigenstate); this could work as a selection criterion in the search of valid Hamiltonians for a given physical system. Interestingly, MM keeps the potential trivial even in $z$ space; it suggests that the map would be closer to the original. However the solutions are far from trivial.

 {In Section~\ref{sec:pmdh} we applied the approach to the context of a one-dimentional double-heterostructure. We considered no external potential but a PDM of parabolic type in the joining region of two different ordinary constant mass values $m_1$ and $m_2$. We found similarities with the case of ordinary quantum mechanics with transitions among constant potentials (e.g. finite potential-well) where bound states and scattering states can be found depending on the particle energy values. We analytically computed the energy eigenstates as well as reflection and transmission coefficients. For the chosen parameters, the different orderings showed very similar results as can be seen in Fig.\ref{fig:10} and Fig.\ref{fig:11}.
%(na verdade, esta sutileza nas diferenças entre os Hamiltonianos foi observada em todos os perfis apresentados ao longo da Seção~\ref{sec:massprof}). 
Reflection and transmission coefficients present oscillatory behaviour,  respectively decreasing with energy and rapidly convergent to zero, in the first case, or growing and rapidly convergent to \(\sqrt{m_2/m_1}\) in the second.}

Our results signal that once the spectrum of a given material is experimentally available one could choose the better arrangement among the collection just examined to model the specific material or heterostructure. Indeed, although the external potential is zero, PDM eigenfunctions are not actual free states but a sort of effective waves in a solid sample. This is precisely the origin of the position dependent mass. These states can be more free or more bounded depending on the specific effective mass and operator ordering chosen. Discrete energy eigenstates would represent effective bound electronic states in a crystal, while continuous eigenstates would represent in-solid conduction carriers. %in a wire
%It is worth noting that for some eigenenergies the probability to find the particle peaks exactly where the mass does; however, for other states it is the opposite. 
%
Our study is expected to be useful for applications in solid-state structures with the addition of an external potential term.
% as well as for the calculation of information entropy \cite{sears:parr:1980}.

\begin{landscape}%\scriptsize
\begin{longtable}{||c||c|m{2.4cm}||c|m{2.5cm}||c|m{2.5cm}||c|m{2.5cm}||c|m{2.9cm}||}
    \caption{Table of the energy spectra and \(\omega_{a,b}\) parameters for each ordering and mass profile (acronyms in parentheses next to the numerical values of \(\omega_{a,b}\) denote the shapes of the effective potentials  {-- EP}: (IB) for infinite barriers, (FB) for finite barriers, (BB) for bottomless barriers, (IW) for infinite wells, (FW) for finite wells and (BW) for bottomless wells).}
    \label{tab:spectra} \\
    \hline \hline
    \multirow{2}{*}{\textbf{Ordering}} & \multicolumn{2}{c||}{\scriptsize\textbf{ {soliton-like}}} & \multicolumn{2}{c||}{{\scriptsize \textbf{Reciprocal  {biquadratic}}}} & \multicolumn{2}{c||}{{\scriptsize \textbf{Reciprocal quadratic }}} & \multicolumn{2}{c||}{\scriptsize\textbf{Exponential}} & \multicolumn{2}{c||}{\scriptsize\textbf{Parabolic}} \\
    \cline{2-11}
     & {\scriptsize \boldmath\(\omega_{a,b}\)}  & {\scriptsize \textbf{Energy spectrum \newline {(EP)}}}  & {\scriptsize \boldmath\(\omega_{a,b}\)} & {\scriptsize \textbf{Energy spectrum \newline {(EP)}}}  &  {\scriptsize \boldmath\(\omega_{a,b}\)}  & {\scriptsize\textbf{Energy spectrum \newline {(EP)}}}  & {\scriptsize \boldmath\(\omega_{a,b}\)} & {\scriptsize \textbf{Energy spectrum \newline {(EP)}}}  & {\scriptsize \boldmath\(\omega_{a,b}\)} & {\scriptsize \textbf{Energy spectrum \newline {(EP)}}} \\
    \hline
    BDD &   $\frac{3}{4}$   &   Discrete, with minimum; (IW)                    &   $2$  &  Discrete, with minimum; (IW)        &   $\frac{1}{4}$   &   Continuous, with minimum; (FB)                                              &   $\frac{3}{4}$   &   Discrete, with minimum; (IW)        &   $\frac{5}{16}$  &   {\scriptsize Depends on $V(x)$ and $M(x)$ outside \(|x|<\ell\)}; (IB)                                                   \\
    \hline
    GW  &   $-\frac{5}{4}$  &   \textit{No spectrum}; (BB)                      &   $-4$ &  \textit{No spectrum}; (BB)          &   $\frac{1}{4}$   &   Continuous, with minimum; (FB)                                              &   $-\frac{5}{4}$  &   \textit{No spectrum}; (BB)          &   $-\frac{7}{16}$ &   \textit{No spectrum}; (BW)                                                                                              \\
    \hline
    ZK  &   $-\frac{1}{4}$  &   {\textit{No spectrum}; (BB)}    &   $0$  &  Discrete, with minimum; ($cons=-1$) &   $-\frac{3}{4}$  &   Continuous, with minimum {and only one bound state}; (FW)   &   $-\frac{1}{4}$  &   Discrete, with minimum; (BB)        &   $-\frac{3}{16}$ &   \textit{No spectrum}; (BW)                                                                                              \\
    \hline
    LK  &   $-\frac{1}{4}$  &  {\textit{No spectrum}; (BB)}    &   $-1$ &  \textit{No spectrum}; (BB)          &   $\frac{1}{4}$   &   Continuous, with minimum; (FB)                                              &   $-\frac{1}{4}$  &   Discrete, with minimum; (BB)        &   $-\frac{1}{16}$ &   \textit{No spectrum}; (BW)                                                                                              \\
    \hline
    MM  &   $0$             &   Discrete, with minimum; ($cons=0$)              &   $0$  &  Discrete, with minimum; ($cons=0$)  &   $0$             &   Continuous, with minimum;  ($cons=0$)                                       &   $0$             &   Discrete, with minimum; ($cons=0$)  &   $0$             &   {\scriptsize Only $\tilde{E}=0$ is possible; still, it depends on  $V(x)$ and $M(x)$ outside \(|x|<\ell\)}; ($cons=0$)    \\
    \hline
    \hline
\end{longtable}
\end{landscape}

\section*{Acknowledgement}

The authors would like to thank Fundo Nacional de Desenvolvimento da Educação do Ministério da Educação (FNDE) for a PET fellowship.

\bibliographystyle{ieeetr}

\bibliography{refpdm}

\end{document}